\documentclass[11pt,a4paper]{article}
\pdfoutput=1
\usepackage[utf8]{inputenc}
\usepackage{bbm}
\usepackage{fixmath}
\usepackage{jcappub}
\newcommand{\mathe}{\mathrm{e}}
\newcommand{\mathi}{\mathrm{i}}
\newcommand{\mathpi}{\pi}
\let\oldre\Re
\let\oldim\Im
\renewcommand{\Re}{\oldre\mathfrak{e}\,}
\renewcommand{\Im}{\oldim\mathfrak{m}\,}

\newcommand{\dalembert}{\square}
\newcommand{\total}{\operatorname{d}\!}
\newcommand{\christoffel}[4][]{\,{}{#1}\Gamma^{#2}_{#3 #4}\,}
\newcommand{\varchristoffel}[4][]{\,{}{#1}\tilde\Gamma^{#2}_{#3 #4}\,}
\newcommand{\eqend}[1]{\,\text{#1}}
\newcommand{\bigo}[1]{\mathcal{O}\left({#1}\right)}

\newcommand{\Ein}{\operatorname{Ein}}
\DeclareMathOperator*{\distlim}{d-lim}
\newcommand{\sgn}{\operatorname{sgn}}
\newcommand{\tr}{\operatorname{tr}}
\newcommand{\abs}[1]{{\left\lvert{#1}\right\rvert}}
\newcommand{\bra}[1]{{\left\langle{#1}\right\rvert}}
\newcommand{\ket}[1]{{\left\lvert{#1}\right\rangle}}
\newcommand{\dotprod}[2]{{\langle{#1}\vert{#2}\rangle}}
\newcommand{\expect}[1]{{\left\langle{#1}\right\rangle}}
\newcommand{\op}[1]{\hat{#1}}
\newcommand{\unitmatrix}{\mathbbm{1}}
\renewcommand{\vec}[1]{\mathbold{#1}}

\title{One-loop gravitational wave spectrum in de Sitter spacetime}
\author[a]{Markus B. Fröb,}
\author[b,c]{Albert Roura}
\author[a]{and Enric Verdaguer}
\affiliation[a]{Departament de Física Fonamental, Institut de Ciències del Cosmos (ICC),\\
Universitat de Barcelona (UB), C/ Martí i Franquès 1, 08028 Barcelona, Spain}
\affiliation[b]{Max-Planck-Institut für Gravitationsphysik (Albert-Einstein-Institut),\\
Am Mühlenberg 1, 14476 Golm, Germany}
\affiliation[c]{Institut für Quantenphysik, Universität Ulm,\\
Albert-Einstein-Allee 11, 89081 Ulm, Germany}
\emailAdd{mfroeb@ffn.ub.edu}
\emailAdd{albert.roura@uni-ulm.de}
\emailAdd{enric.verdaguer@ub.edu}
\abstract{The two-point function for tensor metric perturbations around de Sitter spacetime including one-loop corrections from massless conformally coupled scalar fields is calculated exactly. We work in the Poincaré patch (with spatially flat sections) and employ dimensional regularization for the renormalization process. Unlike previous studies we obtain the result for arbitrary time separations rather than just equal times. Moreover, in contrast to existing results for tensor perturbations, ours is manifestly invariant with respect to the subgroup of de Sitter isometries corresponding to a simultaneous time translation and rescaling of the spatial coordinates. Having selected the right initial state for the interacting theory via an appropriate $\mathi\epsilon$ prescription is crucial for that. Finally, we show that although the two-point function is a well-defined spacetime distribution, the equal-time limit of its spatial Fourier transform is divergent. Therefore, contrary to the well-defined distribution for arbitrary time separations, the power spectrum is strictly speaking ill-defined when loop corrections are included.}
\keywords{cosmological perturbation theory, power spectrum, quantum field theory on curved space}

\begin{document}
\maketitle

\section{Introduction}
\label{introduction}

The amplification of quantum vacuum fluctuations in inflationary models provides a natural mechanism for the generation of primordial cosmological inhomogeneities. This constitutes the key element of the currently most successful paradigm for explaining the observed cosmic microwave background (CMB) anisotropies and large scale structure of the universe. The standard theoretical analysis and calculation of the primordial spectrum relies on the relatively simple procedure of quantizing linear cosmological perturbations, which amounts to a tree-level calculation in an accelerating FLRW background (typically with quasi-exponential expansion and, hence, close to de Sitter spacetime). This framework also predicts quite generically the existence of an approximately scale-invariant background of primordial gravitational waves (tensor perturbations).
The situation, however, becomes much more complex and subtle when quantum effects at nonlinear level are taken into account and loop corrections are considered. In fact, a clear understanding of how significant the effects in this context are is still an unsettled issue.

Tsamis and Woodard proposed that radiative corrections from graviton loops at two-loop order and higher could generate a secular screening of the cosmological constant \cite{tsamis96,tsamis97}. The basic idea is that in a certain gauge free gravitons behave like two massless minimally coupled scalar fields (one for each polarization), and more and more modes get excited as they leave the horizon without being diluted away by the exponential expansion. It is then argued that due to the nonlinear nature of the gravitational interaction, the back-reaction of the attractive gravitational field between the growing number of IR graviton modes slows down the accelerated expansion \cite{tsamis98}.

In addition, there has recently been renewed interest in the possibility that quantum effects in interacting theories could lead to an instability even for massive theories evolving on a fixed de Sitter background \cite{polyakov}. This possibility has been analyzed both at tree level \cite{bros,higuchi09a,higuchi09,alvarez,akhmedov2008,akhmedov10} and including loop corrections \cite{krotov,jatkar}. In this context, the use of Euclidean methods has proven very useful. They provide a natural and direct way of generating the Hartle-Hawking state \cite{jacobson} for the interacting theory, which generalizes the de Sitter-invariant Bunch-Davies vacuum of the free theory to the interacting case. It exhibits de Sitter-invariant quantum correlation functions, which can be obtained by analytic continuation from Euclidean to Lorentzian time, and has been shown to constitute a late-time attractor (defined in a precise sense) for the evolution of generic initial states \cite{marolf2010,marolf2011a,hollands}. Although these studies focused on sufficiently massive theories (depending on the coupling strength), generalizations to massless and very light fields have also been developed \cite{rajaraman,hollandsmassless}.

Our goal is to extend these considerations to the gravitational case. As a first step, we will concentrate on the effects of matter loops (neglecting graviton loops) on the two-point function characterizing the quantum fluctuations of the metric perturbations around a de Sitter background. There are clearly significant differences between the main case considered in the studies mentioned in the previous paragraph, a scalar field with no derivative interaction, and the gravitational case. One is the (partial) derivative character of the gravitational interaction, which improves for instance the IR behavior of massless minimally coupled matter fields in this case. (In addition, perturbative quantum gravity around a given background is power-counting non-renormalizable and should be treated as a low energy effective field theory \cite{burgess,donoghue}, but this aspect is shared by those analyses that considered scalar fields with nonlinear interactions in arbitrary dimensions \cite{higuchi11}.)
A second difference is the existence of a gauge symmetry associated with diffeomorphism invariance and the need to consider appropriate gauge-invariant observables, which is a rather nontrivial aspect even in perturbative quantum gravity \cite{giddings}. Moreover, one should restrict one's attention to ``sufficiently local'' observables that properly characterize the geometrical properties within a region of finite physical size. This point has been crucial in order to construct IR-safe observables in situations which would otherwise lead to divergences in the absence of an IR cut-off \cite{urakawa,gerstenlauer,giddings,senatore12}.
Here we calculate the one-loop correction to the tensorial metric perturbations, which is the key ingredient to obtain such kind of observables. The scalar and vectorial metric perturbations, in contrast, can be directly obtained from the stress tensor correlation function.

Our strategy (mainly for computational simplicity) will be to do the calculation using spatially flat coordinates in the Poincaré patch of de Sitter spacetime. We will employ the closed-time-path (CTP, also known as \emph{in-in}) formalism, which gives the real time evolution of true expectation values and correlation functions rather than transition matrix elements \cite{chou85}. Moreover, an asymptotic initial state selected by an appropriate Wick-rotation prescription for the two CTP branches will be considered. This corresponds to an asymptotic adiabatic vacuum of the interacting theory which is expected to coincide with the Hartle-Hawking state generated by a Euclidean path integral. In fact, the equivalence between such a Euclidean calculation of the correlation functions (followed by an analytic continuation to Lorentzian time) and the \emph{in-in} calculation has been established for massive scalar theories in \cite{higuchi11}. That result, however, cannot be directly extended to the gravitational case because the graviton propagator does not fall-off quickly enough for large spacetime separations. Nevertheless, we expect that as long as one considers gauge-invariant observables sensitive only to geometric properties localized within a region of finite physical size, the two methods will yield equivalent results.

The effects of loops of conformal fields on tensor cosmological perturbations have recently been studied in \cite{wuetal11}, where a result was found which grows as the initial time tends to minus infinity and diverges in that limit. That kind of result together with the small amplitude of the observed CMB anisotropies would impose a bound on the possible number of e-foldings during inflation. It is also incompatible with de Sitter invariance, which implies (as a necessary but not sufficient condition) that there should be a (co-moving) time translation symmetry for spatially flat sections provided that one compares properties at different times involving the same physical distance and momentum scales.

Exact de Sitter invariance will only happen for a very particular choice of initial state corresponding to the de Sitter-invariant vacuum of the interacting theory (rather than the free one): in general results will not be exactly invariant for any other state. However, if such a de Sitter-invariant state exists and is a late-time attractor, in analogy with the results for scalar theories \cite{marolf2011a,hollands} and as suggested by the intuition that regular initial excitations get red-shifted away by the exponential expansion, one expects that for sufficiently regular initial states the result for observables characterizing the local geometry in a finite region should become insensitive to the initial conditions and tend at sufficiently late times (or when the initial time tends to minus infinity) to the de Sitter-invariant result.

These are of course just expectations which should be confirmed by an actual computation. Here we provide an explicit and exact calculation (at one loop in the matter fields) with a result which is well behaved in the limit of asymptotic initial time and respects the translation symmetry (rescaling symmetry for co-moving momenta and conformal times) mentioned above. Our result also confirms the absence of terms involving a logarithmic dependence with respect to the co-moving momentum only. (Such terms, which violate the rescaling symmetry, were originally found in calculations of one-loop corrections to the power spectrum of scalar \cite{weinberg05,chaicherdsakul07} and tensor \cite{aelim09} cosmological perturbations, but were later argued not to be correct in \cite{senatore}, where their absence was explicitly shown for scalar perturbations.) Furthermore, we will explain in detail how the behavior found in \cite{wuetal11} is a consequence of not considering the full contribution of matter fields to the quantum correlation function of the metric perturbations (at one loop).

It is worth emphasizing that we have obtained the correlation function for two arbitrary times rather than just the equal-time limit, as done in previous studies. This is an important point if one wants to check full de Sitter invariance: by considering only equal times, instead, one can only check invariance under a subgroup of the de Sitter group. Moreover, together with an explicit implementation of the renormalization procedure, our exact result for two arbitrary times has enabled us to uncover a previously overlooked subtlety: although the one-loop two-point function is a well-defined spacetime distribution, the power spectrum (the equal-time limit of its spatial Fourier transform) is divergent, even though all usual divergences arising in the calculation have been properly regulated and canceled out by local counterterms.

The paper is organized as follows.
In section~\ref{overview} we briefly review the in-in (or CTP) formalism and the $\mathi \epsilon$ prescription which selects in this context the adiabatic vacuum of the interacting theory as the asymptotic initial state. Next, a general expression for the quantum two-point function at one-loop is derived in this framework. It can be conveniently written in terms of the effective action that results from integrating out the matter fields. Such an effective action is presented in section~\ref{effective_action} for metric perturbations around a FLRW background when integrating out a conformal scalar field. Making use of these tools, in section~\ref{twopoint_func} we compute the two-point function of the tensor perturbations around a de Sitter background for an arbitrary pair of points and including the one-loop correction from a conformal field. Moreover, we give a detailed explanation for the discrepancy of our results as compared to those of \cite{wuetal11}. Finally, we discuss our main findings in section~\ref{discussion}. A number of technical aspects are contained in the appendices. Furthermore, a thorough discussion of the fact that the power spectrum is, strictly speaking, ill defined when including loop corrections is provided in appendix~\ref{appendix_coincidence}.

We use the ``+++'' sign convention of \cite{mtw}, but use only Latin tensor indices which range over space and time. Throughout the paper we work in natural units $c = \hbar = 1$ and take $\kappa^2 = 16 \mathpi G_\text{N}$, where $G_\text{N}$ is Newton's constant.

\section{A general overview of the formalism and calculation}
\label{overview}

\subsection{The \texorpdfstring{``in-in''}{"in-in"} formalism and the \texorpdfstring{$\mathi\epsilon$}{i e} prescription}
\label{overview_inin}

In the quantum field theoretical treatment of scattering problems one usually calculates the transition matrix element of an operator between two states,
\begin{equation}
\bra{\alpha} \op{A} \ket{\beta} \eqend{,}
\end{equation}
where $\ket{\alpha}$ and $\ket{\beta}$ can be taken to be two different \emph{in} and \emph{out} vacuum states, with the particle content of the real states incorporated into the operator $\op{A}$. Up to a (possibly infinite) proportionality constant, one has the path integral representation
\begin{equation}
\label{matrix_element_as_path_integral}
\bra{\text{out}} \op{A} \ket{\text{in}} \sim \int A[\phi] \mathe^{\mathi S[\phi]} \mathcal{D} \phi \eqend{,}
\end{equation}
assuming $\op{A}$ to be time-ordered (which is no problem in practice) and denoting by $\phi$ the field content of the theory, and by $S$ the action.

If $\op{A}$ is a polynomial in the fields (and their derivatives), or can be well approximated by a polynomial, one can calculate its transition matrix elements by adding a classical source $J$ to the action
\begin{equation}
\label{action_with_source}
S \to S + \int J(x) \phi(x) \total^n x \eqend{,}
\end{equation}
where $n$ is the number of spacetime dimensions, and constructing the \emph{generating functional}
\begin{equation}
\label{generating_functional_as_path_integral}
Z[J] = \int \mathe^{\mathi S[\phi, J]} \mathcal{D} \phi \eqend{,}
\end{equation}
in terms of the external source $J$\footnote{In a gravitational context, general covariance of the action \eqref{action_with_source} requires that the source $J(x)$ transforms as a density under general coordinate changes.}. Normalized matrix elements of time-ordered polynomials in the fields are then obtained by functionally differentiating with respect to the source $J$,
\begin{equation}
\frac{\bra{\text{out}} \mathcal{T} \op{\phi}(x_1) \cdots \op{\phi}(x_n) \ket{\text{in}}}{\dotprod{\text{out}}{\text{in}}} = \left[ Z^{-1}[J] \, \frac{\delta}{\mathi \delta J(x_1)} \cdots \frac{\delta}{\mathi \delta J(x_n)} Z[J] \right]_{J = 0} \eqend{,}
\end{equation}
so that we can also display the generating functional as
\begin{equation}
\label{generating_funtional_inout}
Z[J] = \bra{\text{out}} \mathcal{T} \mathe^{\mathi \int J(x) \op{\phi}(x) \total x} \ket{\text{in}} \eqend{.}
\end{equation}

To implement the standard flat-space choice of vacuum in the path integral \eqref{matrix_element_as_path_integral} or \eqref{generating_functional_as_path_integral}, one slightly tilts the time integration contour on the complex plane to include an imaginary part
\begin{equation}
t \to t (1 - \mathi \epsilon)
\end{equation}
with $\epsilon > 0$ (see for instance section 4.2 in \cite{peskinschroeder}). This selects the asymptotic vacuum as the state of lowest energy of the full interacting theory, which includes appropriate correlations between the different fields or even different modes of the same field.\footnote{This procedure selects the ground state of the interacting theory only for a time-independent Hamiltonian. Under the appropriate conditions, however, in the time-dependent case, this prescription can still select an adiabatic vacuum of the theory at early times.} In practice, one calculates the integral from some initial time $t_0$ to some final time $T$, and takes the limits $t_0 \to -\infty$, $T \to \infty$ in a slightly imaginary direction; or, as it is often done, rotate the time axis altogether onto the imaginary axis, which gives rise to \emph{Euclidean} quantum field theory.

In a cosmological setting, one is interested in true correlation functions, which are expectation values of operators rather than transition matrix elements. Moreover, one typically needs to impose initial conditions at early times instead of boundary conditions at both early and late times. Furthermore, as was explicitly shown in \cite{higuchi09,polyakov} for de Sitter space, in an exponentially expanding spacetime ``in-out'' perturbation theory has an infra-red divergence because of the expanding space volume. For all these reasons we are naturally led to consider the ``in-in'' formalism, where one specifies initial conditions at some initial time (in certain cases, such as the exponentially expanding patch of de Sitter with spatially flat sections, one can specify these initial conditions at past infinity and define an asymptotic ``in'' vacuum).

In order to calculate an expectation value using path integrals, one inserts the identity operator as a sum over an orthonormal basis of states at some ``final'' time $T$,
\begin{equation}
\begin{split}
\bra{\text{in}} \op{A} \ket{\text{in}} &= \sum_{\alpha} \dotprod{\text{in}}{\alpha,T} \bra{\alpha,T} \op{A} \ket{\text{in}} = \sum_{\alpha} \left( \dotprod{\alpha,T}{\text{in}} \right)^* \bra{\alpha,T} \op{A} \ket{\text{in}} \\
&\sim \int A[\phi^+] \delta\big[ \phi^+(T) - \alpha \big] \delta\big[ \phi^-(T) - \alpha \big] \mathe^{\mathi S[\phi^+]} \mathe^{- \mathi S[\phi^-]} \mathcal{D} \phi^+ \mathcal{D} \phi^- \mathcal{D} \alpha \eqend{,}
\end{split}
\end{equation}
where the states $\ket{\alpha,T}$ correspond to an orthonormal basis of field-configuration eigenvectors in the Heisenberg picture, such that $\hat{\phi}(T) \ket{\alpha,T} = \alpha \ket{\alpha,T}$. Since we have two path integrals for each degree of freedom, we need two copies of the fields which we have labeled $\phi^+$ and $\phi^-$. Instead of enforcing the separate equality of both fields to $\alpha$ and then integrating over all field configurations $\alpha$ at time $T$, in the following we can directly enforce the equality of the fields $\phi^+$ and $\phi^-$ at that time.

Similarly to the ``in-out'' case, a generating functional for the correlation functions can be introduced, but one needs two sources $J^+$ and $J^-$ corresponding to each one of the two path integrals. In addition, because of the complex conjugation of the second path integral's integrand, which effectively reverses the time integration in the action, the corresponding operator expression is naturally anti-time-ordered, which we denote by the symbol $\widetilde{\mathcal{T}}$. Furthermore we may generalize the expression by considering a density matrix $\op{\rho}$ as the initial state, which for an initially pure state reduces to $\ket{\text{in}}\bra{\text{in}}$. Altogether, the corresponding ``in-in'' generating functional is given by
\begin{equation}
\label{ctp_generating_functional_as_path_integral}
\begin{split}
Z[J^+, J^-] &= \tr\left[ \left( \mathcal{T} \mathe^{\mathi \int J^+(x) \op{\phi}(x) \total^n x} \right) \op{\rho} \left( \widetilde{\mathcal{T}} \mathe^{-\mathi \int J^-(x) \op{\phi}(x) \total^n x} \right) \right] \\
&= \int \delta\big[ \phi^+(T) - \phi^-(T) \big] \delta\big[ \phi^+(t_0) - \phi_0^+ \big] \delta\big[ \phi^-(t_0) - \phi_0^- \big] \bra{\phi^+_0} \op{\rho} \ket{\phi^-_0} \times \\
&\qquad\times \mathe^{\mathi ( S[\phi^+] + \int J^+(x) \phi^+(x) \total x ) - \mathi ( S[\phi^-] + \int J^-(x) \phi^-(x) \total x )} \mathcal{D} \phi^+ \mathcal{D} \phi^- \mathcal{D} \phi_0^+ \mathcal{D} \phi_0^- \eqend{,}
\end{split}
\end{equation}
where $\phi_0^\pm$ are the field configurations at the initial time $t_0$. This time can be finite, although one would need in principle to consider an appropriate dressed state for the interacting theory. In this paper instead we will take $t_0 = -\infty$ and use an appropriate prescription to select the right asymptotic initial state, just as in the case described above.

The expression in equation~\eqref{ctp_generating_functional_as_path_integral} can be interpreted as follows: starting from an initial state described by the density matrix $\op{\rho}$, we evolve it forward in time under the influence of an external source $J^+$ up to some final time $T$, and then evolve backwards in time under the influence of a \emph{different} source $J^-$ (were the two sources the same, we would have clearly achieved nothing). Because of this, the ``in-in'' formalism is also known as the closed-time-path (CTP) formalism, where the usual time integration contour going from $-\infty$ to $+\infty$ is replaced by one going from $-\infty$ to $T$ and turning back to $-\infty$. In fact, one can alternatively write the path integral expression for the CTP generating functional in equation~\eqref{ctp_generating_functional_as_path_integral} in terms of a single copy of the fields by having the time integrals in the action run over this whole CTP contour (and letting the fields and the currents have independent values on both halves of the contour) \cite{cooper94}. Similarly, since Feynman rules can be directly derived from the generating functional, one can show that the Feynman-Stückelberg diagrams for the ``in-in'' formalism can be simply obtained from the ``in-out'' diagrams by evaluating the time integrals associated with the interaction vertices along the full CTP contour.

\begin{figure}[ht]
\begin{center}
\includegraphics{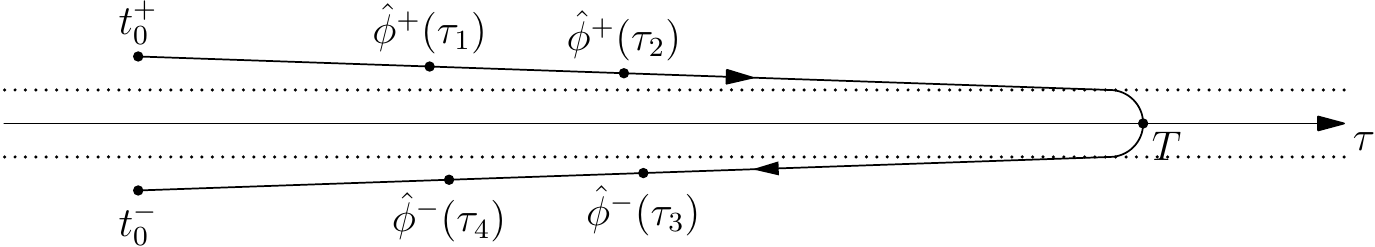}
\end{center}
\caption{The deformed CTP integration contour, together with 4 path-ordered fields. For any permutation of the indices we have $\mathcal{P} \hat{\phi}^\pm(\tau_a) \hat{\phi}^\pm(\tau_b) \hat{\phi}^\pm(\tau_c) \hat{\phi}^\pm(\tau_d) = \hat{\phi}^-(\tau_4) \hat{\phi}^-(\tau_3) \hat{\phi}^+(\tau_2) \hat{\phi}^+(\tau_1)$.
Here, $\hat{\phi}^+(\tau_a)$ and $\hat{\phi}^-(\tau_a)$ simply represent the operator $\hat{\phi}$ evaluated at a time lying, respectively, on the upper and lower branch of the deformed complex contour for time integration, and correspond to the fields $\phi^+$ and $\phi^-$ respectively in the path integrals.}
\label{ctp_path}
\end{figure}

Functionally differentiating the CTP generating functional $Z[J^+, J^-]$ with respect to $\mathi J^+$ and $-\mathi J^-$ one now generates expectation values of ordered polynomials in the fields, which in general are \emph{path-ordered} (denoted by $\mathcal{P}$) along the CTP contour rather than time-ordered. For the $\phi^+$ fields, the path ordering is the standard time ordering, while for the $\phi^-$ fields it is anti-time ordering as discussed above; and $\phi^-$ fields are always to be ordered as if they occurred at a later time than any $\phi^+$ field (see figure~\ref{ctp_path}).

The $\mathi \epsilon$ prescription can also be carried over to the CTP formalism. However, due to the complex conjugation of the integrand in the path integral for the $\phi^-$ fields, we need to take the complex conjugate prescription for them
\begin{equation}
t \to t (1 + \mathi \epsilon) \eqend{.}
\end{equation}
One therefore has to integrate along a contour going from $t_0^+$ to $T$, returning to $t_0^-$ with $t_0^- = (t_0^+)^*$, and taking at the end of the calculation $t_0^+ \to -\infty (1 - \mathi \epsilon)$. The dependence on $T$ disappears in the final result as long as it is larger than all times of interest (i.e., all the times in the arguments of the correlation function one wants to calculate), as required by causality. The deformed contour and the path ordering are shown in figure~\ref{ctp_path}.

Note that this prescription is suitable for selecting the asymptotic initial state in spacetimes where the behavior of the modes for free fields is dominated at past infinity by the same kind of oscillatory behavior as in Minkowski space (factors with a power-law or weaker time dependence are allowed); it then selects the adiabatic vacuum for the interacting theory as $t \to -\infty$. This is the case for the exponentially expanding patch of de Sitter spacetime with spatially flat sections, but would not be appropriate for global de Sitter, which is instead exponentially contracting coming from past infinity.

For further details about the CTP formalism see for instance \cite{chou85,cooper94,jordan86,jordan87,calzettahu1987,camposverdaguer94}.

\subsection{Calculating the two-point function at one-loop order}
\label{overview_calculus}

Taking a mean field approach to quantum gravity, we will quantize the metric perturbations around a semiclassical background including the effects of matter loops. Although perturbative quantum gravity is power-counting nonrenormalizable, it can be consistently studied as an effective field theory (EFT) describing quantum gravitational phenomena with characteristic length scales well above the Planck length $l_\text{p}$ \cite{donoghue,burgess}. In order to do so, one needs to introduce local counterterms in the bare action for the metric and the matter fields with an arbitrary number of derivatives and powers of the curvature, suppressed by the corresponding negative power of the Planck mass (or positive power of the Planck length). The key point is that for phenomena with a characteristic length scale $L$ only a finite number of counterterms needs to be considered to achieve a certain precision, roughly given by $(l_\text{p}/L)^{2n}$ with $n$ being the total number of pairs of derivatives and/or powers of the curvature.

More specifically, we want to calculate the two-point function of the metric perturbations around a fixed background $g^0_{ab}$ (so that the total metric is $g_{ab} = g^0_{ab} + \kappa h_{ab}$) coupled to a scalar field. Therefore, we will consider the following CTP generating functional containing external sources for the metric perturbations $h_{ab}$ (since we are not interested in the matter field correlations, we do not need sources for the matter field $\phi$):
\begin{equation}
\begin{split}
Z[J^+, J^-] &= \int \delta\big[ \phi^+(T) - \phi^-(T) \big] \delta\big[ h^+(T) - h^-(T) \big] \delta\big[ h^\pm(t_0) - h^\pm_0 \big] \delta\big[ \phi^\pm(t_0) - \phi^\pm_0 \big] \times \\
&\qquad\times \bra{\phi^+_0, h^+_0} \op{\rho} \ket{\phi^-_0, h^-_0} \mathe^{\mathi ( S[h^+, \phi^+] + J^+ h^+ ) - \mathi ( S[h^-, \phi^-] + J^- h^- )} \mathcal{D} \phi^\pm \mathcal{D} h^\pm \mathcal{D} \phi_0^\pm \mathcal{D} h_0^\pm \eqend{,}
\end{split}
\end{equation}
where $S[h, \phi]$ denotes the total action for metric perturbations and matter fields, $\phi^\pm_0$ and $h^\pm_0$ correspond to the field configurations at the initial time $t_0$, and $\hat{\rho}$ is the density matrix specifying the initial state. Note that for simplicity we adopt throughout this subsection a shorthand notation with tensor indices, spatial dependence of the fields and spacetime integrals omitted. Our calculation will be performed in two steps: first, we explicitly integrate over the matter fields $\phi^\pm$, and next we perform the remaining path integrals for the metric perturbations $h^\pm$. It is easier to proceed if one assumes that the initial density matrix factorizes so that $\bra{\phi^+_0, h^+_0} \op{\rho} \ket{\phi^-_0, h^-_0} = \bra{\phi^+_0} \op{\rho}_\phi \ket{\phi^-_0} \bra{h^+_0} \op{\rho}_h \ket{h^-_0}$. We are actually interested in states of the interacting theory which involve correlations (entanglement) between the matter fields and the metric perturbations, for which the density matrix does not factorize. However, if one employs the method described in the previous subsection to select the adiabatic vacuum of the interacting theory, one can consider a factorized initial state involving the product of the Bunch-Davies vacua for the matter fields and the metric perturbations both treated as free fields, integrate out the matter fields more easily and only at the final stage deform the CTP integration contour (as shown in figure~\ref{ctp_path}) for the vertices appearing in the Feynman-Stückelberg diagrams associated with the perturbative calculation of the path integral of the metric perturbations.
In doing so we will neglect contributions from graviton loops. This can be formally implemented in a natural way by considering a large $N$ expansion for $N$ matter fields interacting with the gravitational field \cite{hrv_orderredux,hartlehorowitz}; for instance, the lowest-order contributions to the connected two-point function of the metric perturbations are of order $1/N$, whereas any contribution including graviton loops is suppressed by higher powers of $1/N$.

After integrating out the matter fields, which amounts to a one-loop calculation, one is left with the following expression for the CTP generating functional (dropping an infinite proportionality constant, which anyway will cancel in the calculation of the correlation function):
\begin{equation}
\label{generating_functional_effective}
\begin{split}
Z[J^+, J^-] &= \int \delta\big[ h^+(T) - h^-(T) \big] \delta\big[ h^\pm(t_0) - h^\pm_0 \big] \times \\
&\qquad\times \bra{h^+_0} \op{\rho}_h \ket{h^-_0} \mathe^{\mathi ( S_\text{G}[h^+] + J^+ h^+ ) - \mathi ( S_\text{G}[h^-] + J^- h^- ) + \mathi \Sigma[h^+,h^-]} \mathcal{D} h^\pm \mathcal{D} h_0^\pm \eqend{,}
\end{split}
\end{equation}
where $S_\text{G}[h]$ is the gravitational part of the original bare action $S[h,\phi]$, i.e. all the terms depending only on the metric, and $\Sigma$ is the result of the functional integration over the matter fields, truncated at second order in $h$.
In the course of ``integrating out'' the matter fields $\phi$, the ultraviolet (UV) divergences associated with matter loops which appear in $\Sigma$ must be regularized and appropriate counterterms must be included in the bare gravitational action $S_\text{G}$. To leading order in $1/N$, it is only necessary to introduce counterterms in $S_\text{G}[h]$ at most quadratic in the curvature\footnote{Since perturbative quantum gravity around a given background is power-counting non-renormalizable, for each additional term in the $1/N$ expansion one would need to include counterterms with higher and higher powers of the curvature and number of covariant derivatives in $S_\text{G}[h]$. Moreover, in that case one would also need to include in the full bare action $S[h,\phi]$ counterterms coupling $h$ and $\phi$, and with a higher and higher number of covariant derivatives.}, so that the bare gravitational action is schematically given by
\begin{equation}
\label{bare_grav_action}
S_\text{G}[g] = \frac{N}{\bar{\kappa}^2} \left[ \alpha_0 \int (R-2\Lambda) \sqrt{-g} \total^n x + \alpha_2 \bar{\kappa}^2 \int R^2 \sqrt{-g} \total^n x + \mathcal{O}\left(\bar{\kappa}^4\right) \right] \eqend{,}
\end{equation}
where $\alpha_i$ are dimensionless bare parameters and $R^2$ has been used as a shorthand for terms quadratic in the Riemann tensor with all possible contractions. We have introduced a rescaled gravitational coupling constant
\begin{equation}
\bar{\kappa}^2 = N \kappa^2 \eqend{,}
\end{equation}
which helps to organize conveniently the calculation when using a $1/N$ expansion.\footnote{Note that having a fundamental length-scale $\bar{\kappa}$ characterizing the breakdown of the low-energy gravitational EFT which is much larger than the effective Planck length $l_\mathrm{p} = \sqrt{16\pi} \kappa$ determined at low energies can happen naturally in braneworld models with warped extra dimensions \cite{rs2} or models with a large number of fields \cite{dvali}. In fact, in some cases both possibilities can be understood as equivalent descriptions within the framework of the AdS/CFT correspondence \cite{gubser}.} Having done so, the divergent terms cancel out in the sum
\begin{equation}
\label{effective_action_definition}
S_\text{eff}[h^+, h^-] = S_\text{G}[h^+] - S_\text{G}[h^-] + \Sigma[h^+, h^-] \eqend{,}
\end{equation}
and the regulator can be taken to its physical value. Note that $\Sigma$ provides an effective propagator for $h$ which includes the interaction with $\phi$, so that $S_\text{eff}$ can be called a \emph{CTP effective action}. 

From now on we will only consider $S_\text{eff}[h^+,h^-]$ through quadratic order in the metric perturbations since higher-order terms do not contribute to the generating functional and the connected correlation functions to leading order in $1/N$, as mentioned above. To analyze more closely the different contributions to $S_\text{eff}$, let us expand it in powers of $h$,
\begin{equation}
\label{effective_action_expansion_h}
\begin{split}
S_\text{eff}[h^+,h^-] &= S^{(0)}_\text{G}[h^+] + S^{(1)}_\text{G}[h^+] + S^{(2)}_\text{G}[h^+] - S^{(0)}_\text{G}[h^-] - S^{(1)}_\text{G}[h^-] - S^{(2)}_\text{G}[h^-] \\
&\qquad+ \Sigma^{(1)}[h^+, h^-] + \Sigma^{(2)}[h^+, h^-] \eqend{,}
\end{split}
\end{equation}
where the superscript denotes how many powers of $h$ appear. The sum of all the first order terms corresponds to integrating $h^+$ and $h^-$ with the functional derivative of the CTP effective action for the matter fields in a semiclassical background with the metric of the two branches equal to $g^0_{ab}$ after differentiation. When equated to zero, such a functional derivative corresponds to the semiclassical Einstein equation which governs the quantum back-reaction of the matter fields on the mean background geometry \cite{wald,flanagan96}
\begin{equation}
\label{semiclassical_Einstein}
G_{ab}\left[ g^0 \right] - \Lambda g^0_{ab} = \frac{\kappa^2}{2} \expect{ \hat{T}_{ab}\left[ g^0 \right] }_\text{ren} \eqend{,}
\end{equation}
where the stress tensor expectation value comes from the functional derivative of $\Sigma$ with respect to $h_{ab}$ and here also includes any finite contributions from the counterterms in $S_\text{G}$ other than the Einstein tensor or the cosmological constant. We will always consider a background $g^0_{ab}$ which is a solution of the semiclassical equation~\eqref{semiclassical_Einstein}. Therefore, the sum of the first-order terms on the right-hand side of equation~\eqref{effective_action_expansion_h} will vanish. Such a choice will also guarantee that after the functional integration there will be no term linear in $J$ in the exponent of the CTP generating functional, which will be purely quadratic at the order at which we are working.

Furthermore, the zeroth-order terms in equation~\eqref{effective_action_expansion_h} can be factored out of the path integral~\eqref{generating_functional_effective} and give a factor independent of $J$ which does not contribute to the correlation functions, obtained by functionally differentiating with respect to $J$. Disregarding them we are, thus, left with a purely quadratic expression for $S_\text{eff}$. Given an exponent quadratic in the perturbations $h^+$ and $h^-$, one can easily perform the functional integration in equation~\eqref{generating_functional_effective}. In order to do so, it is convenient to adopt a matrix formulation with  $h_A = (h^+, h^-)$, $J_A =\nolinebreak[4](J^+, -J^-)$ (note the minus sign), $A_{MN} = \begin{pmatrix} A^{++} & A^{+-} \\ A^{-+} & A^{--} \end{pmatrix}$ and sum over repeated capital indices regardless of their position. Expanding now the effective action $S_\text{eff}$ in powers of $\kappa$,
\begin{equation}
\label{effective_action_expansion_kappa}
S_\text{eff}[h^+,h^-] = S^{(2)}_\text{G}[h^+] - S^{(2)}_\text{G}[h^-] + \Sigma^{(2)}[h^+, h^-] = S_0[h^+,h^-] + \kappa^2 S_2[h^+,h^-] \eqend{,}
\end{equation}
we can write it as
\begin{equation}
S_\text{eff}[h^+, h^-] = \frac{1}{2} h_M A_{MN} h_N = \frac{1}{2} h_M (A^0_{MN} + \kappa^2 V_{MN}) h_N \eqend{,}
\end{equation}
where
\begin{equation}
A^0_{MN} = \begin{pmatrix} \frac{\delta^2 S_0[h^+]}{\delta h^+ \delta h^+} & 0 \\ 0 & - \frac{\delta^2 S_0[h^-]}{\delta h^- \delta h^-} \end{pmatrix}
\end{equation}
is the free theory differential operator acting on $h$, and
\begin{equation}
\label{perturbation_v_definition}
V = \begin{pmatrix} \frac{\delta^2}{\delta h^+ \delta h^+} & \frac{\delta^2}{\delta h^+ \delta h^-} \\ \frac{\delta^2}{\delta h^- \delta h^+} & \frac{\delta^2}{\delta h^- \delta h^-} \end{pmatrix} S_2[h^+, h^-]
\end{equation}
is the part arising from the interaction with the matter fields $\phi$. (Note that $S_2$ also includes any counterterms proportional to the terms in the Einstein-Hilbert action which can appear for massive fields or when using Pauli-Villars regularization.) The functional integration in equation~\eqref{generating_functional_effective} leads to
\begin{equation}
Z[J^+, J^-] \sim \int \delta\big[ h^+(T) - h^-(T) \big] \mathe^{\frac{\mathi}{2} h_M A_{MN} h_N + \mathi J_N h_N} \mathcal{D} h \sim \mathe^{- \frac{\mathi}{2} J_M G_{MN} J_N} \eqend{,}
\end{equation}
where $G$ is the inverse operator of $A$, i.e.
\begin{equation}
\label{propagator_definition}
\int A_{AM}(x, y) G_{MB}(y, x') \total^4 y = \delta^4(x-x') \delta_{AB} \eqend{.}
\end{equation}

\begin{figure}[ht]
\begin{center}
\includegraphics[width=\textwidth]{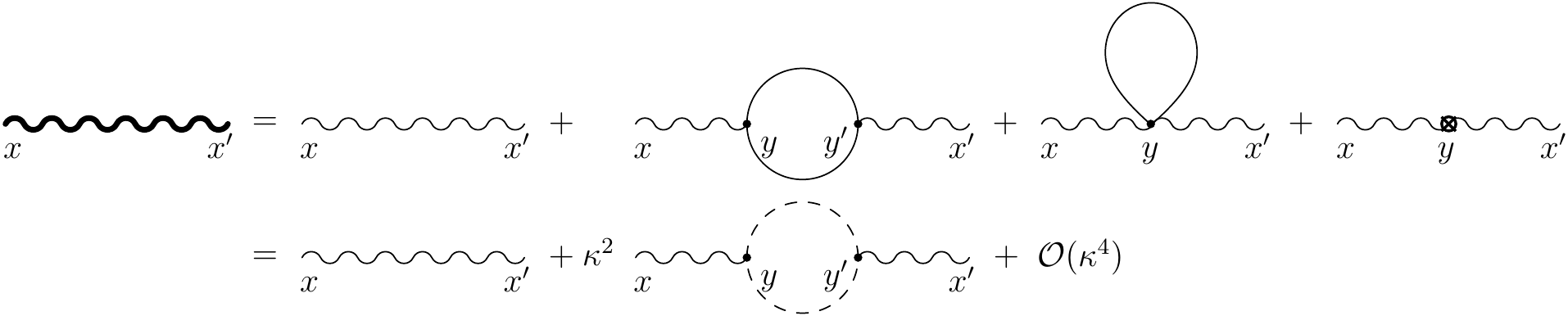}
\end{center}
\caption{The Feynman-Stückelberg diagrammatic expansion corresponding to our calculation. The wiggly lines are gravitons and the straight lines are scalars. Summing the contribution of the loop, the tadpole (which vanishes in dimensional regularization for the massless, conformally coupled scalar in spatially flat FLRW backgrounds) and the counterterms, we obtain the (finite) dashed loop corresponding to $V$ \eqref{perturbation_v_definition}.}
\label{feynman_diagram}
\end{figure}

Functionally differentiating the generating functional twice with respect to the source $J$ (remember that $J_M = (J^+, - J^-)$), we get
\begin{equation}
\label{propagator_functional_derivative}
\bra{\text{in}} \mathcal{P} h_A(x) h_B(x') \ket{\text{in}} = \left[ Z^{-1}[J^+, J^-] \frac{\delta^2 Z[J^+, J^-]}{\mathi \delta J_A(x) \, \mathi \delta J_B(x')} \right]_{J = 0} = \mathi G_{AB}(x, x') \eqend{.}
\end{equation}
Specializing to the four possible values of the index pair $AB$, we get
\begin{align*}
G_{++}(x, x') &= - \mathi \bra{\text{in}} \mathcal{T} h(x) h(x') \ket{\text{in}} = G_\text{F}(x, x') &\!\!\!\!\!\!\!\!\text{(the Feynman propagator)} \\
G_{--}(x, x') &= - \mathi \bra{\text{in}} \widetilde{\mathcal{T}} h(x) h(x') \ket{\text{in}} = G_\text{D}(x, x') &\!\!\!\!\!\!\!\!\text{(the Dyson propagator)}& \\
G_{-+}(x, x') &= - \mathi \bra{\text{in}} h(x) h(x') \ket{\text{in}} = G^+(x, x') &\!\!\!\!\!\!\!\!\text{(the positive Wightman function)}& \\
G_{+-}(x, x') &= - \mathi \bra{\text{in}} h(x') h(x) \ket{\text{in}} = G^-(x, x') &\!\!\!\!\!\!\!\!\text{(the negative Wightman function)}&\footnotemark \eqend{.}
\end{align*}
\footnotetext{Note that while the name \emph{Wightman} is attached to those functions regardless of the order of perturbation theory, one usually uses \emph{Feynman} and \emph{Dyson propagator} only for the tree level result, while in higher orders they are known as time-ordered and anti-time-ordered or $\tau$ functions.}
In general, however, we cannot solve equation~\eqref{propagator_definition} exactly, and must use perturbation theory. Adopting a condensed matrix notation, in which equation~\eqref{propagator_definition} reads $A G = \unitmatrix$, and assuming that we can invert $A^0$ exactly to get the propagator $G^0$
\begin{equation}
A^0 G^0 = \unitmatrix \eqend{,}
\end{equation}
we get from equation~\eqref{propagator_definition} (with appropriate boundary conditions) the exact recursion relation for $G$,
\begin{equation}
G = G^0 - \kappa^2 G^0 V G \eqend{.}
\end{equation}
Since $\kappa^2$ is small, we can solve perturbatively this recursion for $G$ by iterating and truncating
\begin{equation}
G = G^0 - \kappa^2 G^0 V G^0 + \bigo{\kappa^4} \eqend{,}
\end{equation}
which fully written reads
\begin{equation}
\label{propagator_first_order_integral}
G_{AB}(x, x') = G^0_{AB}(x, x') - \kappa^2 \iint G^0_{AM}(x, y) V_{MN}(y, y') G^0_{NB}(y', x') \total^4 y' \total^4 y + \bigo{\kappa^4} \eqend{.}
\end{equation}
This is shown in terms of Feynman-Stückelberg diagrams in figure \ref{feynman_diagram}.

Since $V$ is the second variation \eqref{perturbation_v_definition} of the $S_2[h^+, h^-]$ part of the effective action \eqref{effective_action_expansion_kappa}, the integrand in \eqref{propagator_first_order_integral} is given effectively by $S_2[h^+, h^-]$ with the fields $h$ replaced by the zero-order propagators $G^0$. We will calculate exactly this double integral, adding, of course, the tensor structure which we have ignored until now.

It is worth mentioning that in the stochastic gravity framework~\cite{stochasticgravity,martin99,martin00,hrv04,hvcqg03}, one now rewrites the imaginary part of $S_2[h^+,h^-]$ using an auxiliary stochastic field $\xi$ to arrive at a real \emph{stochastic effective action}, from which manifestly real and causal equations of motion for the perturbation $h$ can be derived. An alternative expression for $G_{AB}$ can then be given as a sum of two terms called ``intrinsic'' and ``induced'' fluctuations, where the ``induced'' fluctuations are a direct consequence of the stochastic source. It is then important to consider the contributions of both terms, since otherwise secular terms occur in the two-point function for the metric perturbations. An explanation of this fact is given in section~\ref{secular_terms}.

However, in order to implement correctly the prescription for the adiabatic ``in'' vacuum in the interacting theory, it is more convenient to work directly with equation~\eqref{propagator_first_order_integral}, as we will do here.

\section{The effective action for metric perturbations around a spatially flat FLRW background}
\label{effective_action}

We consider $N$ massless and conformally coupled free scalar fields interacting with the metric perturbations around a spatially flat Friedmann-Lemaître-Robertson-Walker (FLRW) background driven by a cosmological constant. As described in the previous section, since we are interested in calculating the correlation functions of the metric perturbations to leading order in $1/N$, in addition to the Einstein-Hilbert action for gravity with a cosmological constant we only need to include counterterms quadratic in the curvature for renormalization,
\begin{equation}
\label{EHaction}
\begin{split}
S[\tilde{g}, \phi] &= \frac{N}{\bar{\kappa}^2} \int \left( \tilde{R} - 2 \Lambda \right) \sqrt{-\tilde{g}} \total^n x + N a_1 \int \left( \tilde{R}^{abcd} \tilde{R}_{abcd} - \tilde{R}^{ab} \tilde{R}_{ab} \right) \sqrt{-\tilde{g}} \total^n x \\
&\quad+ N a_2 \int \tilde{R}^2 \sqrt{-\tilde{g}} \total^n x - \frac{1}{2} \sum_{k=1}^N \int \left[ \left( \tilde{\nabla}^m \phi_k \right) \left( \tilde{\nabla}_m \phi_k \right) + \xi(n) \tilde{R} \phi_k^2 \right] \sqrt{-\tilde{g}} \total^n x \eqend{.}
\end{split}
\end{equation}
Note that we are considering this action in $n$ dimensions so that we can use dimensional regularization, and that all parameters are bare parameters, except for $\xi(n) = (n-2)/[4(n-\nolinebreak[4]1)]$ which depends on $n$ such as to make the scalar field action conformally invariant in all dimensions. In the large $N$ limit and after the rescaling of Newton's constant $\bar{\kappa}^2 = N \kappa^2$, graviton loops are suppressed by higher powers of $1/N$ with respect to matter loops, and one can neglect them at leading order. The physical metric $\tilde{g}_{ab}$ is conformal to an almost flat metric~$g_{ab}$
\begin{equation}
\tilde{g}_{ab} = a^2(\eta) g_{ab} = a^2(\eta) ( \eta_{ab} + \kappa h_{ab} ) \eqend{,}
\end{equation}
with $\eta$ denoting the conformal time. Neglecting graviton loops then amounts to truncating the effective action $S_\text{eff}$, as given by equation~\eqref{effective_action_definition}, to second order in the metric perturbation~$h_{ab}$.

The functional integral over the matter fields was calculated in \cite{camposverdaguer94,camposverdaguer96} for the case of a single scalar field ($N=1$). The basic idea is to use the conformal transformation formulas given in appendix~\ref{appendix_conformal} together with the conformal invariance of the classical action for the conformal fields to rewrite equation~\eqref{EHaction} in terms of the metric $g_{ab}$ for small perturbations around flat space. One can then perform the path integral for the scalar fields in flat space treating perturbatively the interaction with the metric perturbations. In doing so, UV divergences arise which are canceled by an appropriate choice of the bare parameter (in dimensional regularization)
\begin{equation}
N a_1 = N a_1^\text{ren}(\mu) + \alpha \left[ (n-4)^{-1} + \ln \mu + \bigo{n-4} \right] \eqend{,}
\end{equation}
where we have introduced the dimensionless constant
\begin{equation}
\alpha = \frac{N}{2880 \mathpi^2} \eqend{.}
\end{equation}
In contrast, the second bare parameter, $N a_2 = (\alpha-\beta)/12$, does not get renormalized, which is a peculiarity of the conformal case~\cite{birrelldavies,camposverdaguer94}. Taking all this into account, the final result for $S_\text{eff}$ obtained in \cite{camposverdaguer94}, which is valid to quadratic order in the metric perturbations, is given by
\begin{equation}
\label{effective_action_in_a_g}
\begin{split}
S_\text{eff}[\tilde{g}^\pm_{ab}] &= S_\text{G}^\text{ren}[a, g^+_{ab}] - S_\text{G}^\text{ren}[a, g^-_{ab}] + 3 \alpha \iint C^+_{abmn}(x) C^{-abmn}(y) K(x-y) \total^4 x \total^4 y \\
&\quad+ \frac{3}{2} \alpha \iint C^+_{abmn}(x) C^{+abmn}(y) K^+(x-y; \mu) \total^4 x \total^4 y \\
&\quad- \frac{3}{2} \alpha \iint C^-_{abmn}(x) C^{-abmn}(y) K^-(x-y; \mu) \total^4 x \total^4 y \\
\end{split}
\end{equation}
with
\begin{equation}
\begin{split}
S_\text{G}^\text{ren}[a, g_{ab}] &= \frac{1}{\kappa^2} \int \left( a^2 R - 6 a \nabla^m \nabla_m a - 2 \Lambda a^4 \right) \sqrt{-g} \total^4 x \\
&\quad+ \alpha \int \left( R^{abcd} R_{abcd} - R^{mn} R_{mn} \right) \ln a \sqrt{-g} \total^4 x \\
&\quad- \frac{\beta}{12} \int \left( R - 6 a^{-1} \nabla^m \nabla_m a \right)^2 \sqrt{-g} \total^4 x \\
&\quad+ 2 \alpha \int G^{mn} a^{-2} \left( \nabla_m a \right) \left( \nabla_n a \right) \sqrt{-g} \total^4 x \\
&\quad+ \alpha \int a^{-4} \left( 2 a \nabla^m \nabla_m a - \left( \nabla^m a \right) \left( \nabla_m a \right) \right) \left( \nabla^n a \right) \left( \nabla_n a \right) \sqrt{-g} \total^4 x \eqend{,}
\end{split}
\end{equation}
where the divergences appearing in $\Sigma$ and those of the bare parameters have canceled out and the remaining parameters are the renormalized ones. Moreover, we have used the identity
\begin{equation}
\label{gauss_bonnet}
\int \left( R^{abcd} R_{abcd} - R^{mn} R_{mn} \right) \sqrt{-g} \total^4 x = \frac{3}{2} \int C^{abcd} C_{abcd} \sqrt{-g} \total^4 x \eqend{,}
\end{equation}
which is valid in four dimensions and follows from the Gauß-Bonnet theorem and the definition of the Weyl tensor in terms of the Riemann and Ricci tensors \eqref{weyl}. The kernels $K$ are given by their Fourier transforms as
\begin{equation}
\label{kernels_def}
\begin{split}
K(x) &= - \mathi \mathpi \int \Theta(-p^2) \Theta(-p^0) \mathe^{\mathi p x} \frac{\total^4 p}{(2\mathpi)^4} \\
K^\pm(x; \mu) &= \frac{1}{2} \int \left[ - \ln \abs{\frac{p^2}{\mu^2}} \pm \mathi \mathpi \Theta(-p^2) \right] \mathe^{\mathi p x} \frac{\total^4 p}{(2\mathpi)^4} \eqend{,}
\end{split}
\end{equation}
where $\mu$ is the renormalization scale. It should be stressed that the dependence on $\mu$ of these kernels and of the renormalized parameters cancels out, so that $S_\text{eff}$ is invariant under the renormalization group. More specifically, we have
\begin{equation}
\label{renormalized_parameter}
N a_1^\text{ren}(\mu) = N a_1^\text{ren}(\mu_0) - \alpha \ln \left( \frac{\mu}{\mu_0} \right)
\end{equation}
for a fixed scale $\mu_0$, and one can see that the logarithmic dependence on $\mu$ of the local terms quadratic in the Weyl tensor exactly matches that of the kernels $K^\pm(x; \mu)$. Given any value of $a_1^\text{ren}(\mu_0)$, one can always choose a renormalization scale $\bar{\mu}$ such that $a_1^\text{ren}(\bar{\mu}) = 0$. This is the choice that was made in~\cite{camposverdaguer94}, and we will also employ it throughout our calculation to deal with slightly more compact expressions. However, once we obtain the final result in the next section, we will specify the small changes needed so that the result is valid for arbitrary $\mu$. On the other hand, as mentioned above, the parameter $a_2$ does not get divergent contributions for conformal fields and, thus, no dependence on $\mu$. In this case the functional integration of the scalar fields just gives a finite contribution, $-\alpha/12$, to the coefficient of the term with the square of the Ricci scalar. To take that into account, the new parameter $\beta = \alpha - 12 N a_2$ has been introduced in equation~\eqref{effective_action_in_a_g}. In~\cite{camposverdaguer94} the particular choice $\beta = \alpha$, corresponding to $a_2 = 0$, was made.

This effective action $S_\text{eff}$ in equation~\eqref{effective_action_in_a_g} should be understood as truncated to second order in $h_{ab}$, the perturbations of the metric $g_{ab}$ around flat space. The expansion of the effective action up to that order can be obtained using the expansions around flat space given in appendix~\ref{appendix_metric}.

In particular, from equations~\eqref{christoffel} -- \eqref{weyl} we have
\begin{equation}
\begin{split}
C_{abcd} &= \frac{\kappa}{3} T_{abcdmnpq} \partial^m \partial^n h^{pq} + \bigo{\kappa^2} \\
T_{abcdmnpq} &= 6 \eta_{p[a} \eta_{b][c} \eta_{d][n} \eta_{q]m} - 6 \eta_{m[a} \eta_{b][c} \eta_{d][n} \eta_{q]p} - 6 \eta_{m[a} \eta_{b]p} \eta_{n[c} \eta_{d]q} + 2 \eta_{a[c} \eta_{d]b} \eta_{m[p} \eta_{n]q} \eqend{.}
\end{split}
\end{equation}
Taking into account that the scale factor $a$ only depends on the conformal time $\eta$ (and denoting the derivative with respect to it by a prime), we then get
\begin{equation}
\label{general_action_in_a_hmn}
\begin{split}
S_\text{eff}[\tilde{g}^\pm_{ab}] &= S_\text{G}^\text{ren}[\tilde{g}^+_{ab}] - S_\text{G}^\text{ren}[\tilde{g}^-_{ab}] \\
&\quad+ \frac{\alpha}{3} \kappa^2 {T_{abcd}}^{mnpq} T^{abcdrskl} \iint \left[ \partial_m \partial_n h^+_{pq}(x) \right] \left[ \partial_r \partial_s h^-_{kl}(x') \right] K(x-x') \total^4 x \total^4 x' \\
&\quad+ \frac{\alpha}{6} \kappa^2 {T_{abcd}}^{mnpq} T^{abcdrskl} \iint \left[ \partial_m \partial_n h^+_{pq}(x) \right] \left[ \partial_r \partial_s h^+_{kl}(x') \right] K^+(x-x'; \mu) \total^4 x \total^4 x' \\
&\quad- \frac{\alpha}{6} \kappa^2 {T_{abcd}}^{mnpq} T^{abcdrskl} \iint \left[ \partial_m \partial_n h^-_{pq}(x) \right] \left[ \partial_r \partial_s h^-_{kl}(x') \right] K^-(x-x'; \mu) \total^4 x \total^4 x' \\
S_\text{G}^\text{ren}[\tilde{g}_{ab}] &= S^{(0)}_\text{G}[a] + S^{(1)}_\text{G}[a, h_{ab}] + S^{(2)}_\text{G}[a, h_{ab}] \eqend{,}
\end{split}
\end{equation}
where
\begin{equation}
\label{action_grav_field_in_a_hmn}
\begin{split}
S^{(0)}_\text{G}[a] &= \frac{1}{\kappa^2} \int \left[ 6 a a'' - 2 \Lambda a^4 + \alpha \kappa^2 a^{-4} \left( 2 a (a')^2 a'' - (a')^4 \right) - 3 \beta \kappa^2 a^{-2} (a'')^2 \right] \total^4 x \\
S^{(1)}_\text{G}[a, h_{ab}] &= \frac{1}{\kappa} \int \left[ 2 a a'' - (a')^2 - \Lambda a^4 + \frac{1}{2} \alpha \kappa^2 a^{-4} \left( 5 (a')^4 - 4 a (a')^2 a'' \right) \right] h \total^4 x \\
&\quad- \frac{1}{2} \beta \kappa \int a^{-3} \left( 2 a^2 a^{(4)} - 10 a a' a''' - 5 a (a'')^2 + 16 (a')^2 a'' \right) h \total^4 x \\
&\quad+ \frac{2}{\kappa} \int \left[ a a'' - 2 (a')^2 - \alpha \kappa^2 a^{-4} \left( a (a')^2 a'' - 2 (a')^4 \right) \right] h^{00} \total^4 x \\
&\quad- \beta \kappa \int a^{-3} \left( a^2 a^{(4)} - 8 a a' a''' - a (a'')^2 + 14 (a')^2 a'' \right) h^{00} \total^4 x
\end{split}
\end{equation}
and $S^{(2)}_\text{G}[a, h_{ab}]$ contains all the terms which are quadratic in the metric perturbation $h_{ab}$.

\subsection{Semiclassical de Sitter background}
\label{desitter_bg}

As explained in section~\ref{overview_calculus}, we will consider a background metric $g^0_{ab}$ which satisfies the semiclassical Einstein equation~\eqref{semiclassical_Einstein}. Assuming a FLRW background of the form $a^2(\eta) \eta_{ab}$, the dynamics for the scale factor $a(\eta)$ is entirely determined by the Friedmann equation, which corresponds to the $00$ component of the semiclassical equation. It can be immediately obtained by functionally differentiating with respect to $h^{00}$ the expression for $S^{(1)}_\text{G}$ in equation~\eqref{action_grav_field_in_a_hmn}. Taking into account that $\delta h(x)/\delta h^{00}(y) = -\delta^{(4)} (x-y)$, we get
\begin{equation}
\label{equation_scalefactor}
6 (a')^2 - 2 \Lambda a^4 = 3 \alpha \kappa^2 a^{-4} (a')^4 + 3 \beta \kappa^2 a^{-3} \left[ 2 a a' a''' - a (a'')^2 - 4 (a')^2 a'' \right] \eqend{.}
\end{equation}
The quantum corrections on the right-hand side of equation~\eqref{equation_scalefactor} involve higher derivative terms, which lead to spurious solutions that lie beyond the domain of validity of the EFT expansion~\cite{simon,flanagan96}. A suitable way of dealing with that problem is the order reduction method~\cite{parker93}. The basic idea, applied to this case, is to start with the equation $6 (a')^2 - 2 \Lambda a^4 = \bigo{\kappa^2}$ and write $a'$ in terms of $a$ up to terms of order $\kappa^2$, which is $a' = (2 \Lambda/3) a^2 + \bigo{\kappa^2}$. Differentiating this expression and proceeding recursively, one can express higher-order derivatives such as $a''$ and $a'''$ in terms of $a$ up order $\kappa^2$. Next, one substitutes these results into the right-hand side of equation~\eqref{equation_scalefactor} and obtains an equation without higher-order derivative terms, but equivalent to equation~\eqref{equation_scalefactor} through order $\kappa^2$, which is anyway the order up to which that equation is valid within the EFT approach to perturbative quantum gravity. When doing so, the terms proportional to $\beta$ cancel out. This was to be expected because they correspond to the trace of the functional derivative with respect to the metric of the term quadratic in the Ricci scalar, which is proportional to $\dalembert R$, whereas the equation that we are substituting implies that $R=4\Lambda$, which is a constant. Taking all that into account, after employing the order reduction method we get
\begin{equation}
3 (a')^2 - \Lambda \left( 1 + \frac{1}{6} \alpha \kappa^2 \Lambda \right) a^4 = \bigo{\kappa^4} \eqend{.}
\end{equation}
We can, therefore, define an effective cosmological constant
\begin{equation}
\label{def_lambda_eff_one}
\Lambda_\text{eff} = \Lambda \left( 1 + \frac{1}{6} \alpha \kappa^2 \Lambda \right) \eqend{,}
\end{equation}
which includes a small constant shift due to the quantum back-reaction of the matter fields on the background geometry. The scale factor is then given by
\begin{equation}
\label{scale_factor}
a(\eta) = - \frac{1}{H \eta}
\end{equation}
with the Hubble parameter $H = \sqrt{\Lambda_\text{eff}/3}$. This corresponds to de Sitter spacetime, more precisely to the part which can be covered with a foliation involving flat spatial sections (the Poincaré patch).

We hereby rederive the well-known fact (see for instance~\cite{dowker,starobinsky80,wada83}) that one can have self-consistent solutions of equation~\eqref{semiclassical_Einstein} which correspond to de Sitter spacetime with a modified cosmological constant, since, due to de Sitter invariance, the renormalized expectation value of the stress tensor for the Bunch-Davies vacuum is a cosmological-constant-like term proportional to the metric.

\subsection{Tensor perturbations}
\label{tensor_perturb}

Concerning the metric perturbations around the mean background, we will from now on restrict to tensor perturbations (scalar and vector perturbations will be discussed in~\cite{frv2012a}), which fulfill the conditions
\begin{equation}
\label{tensor_gauge_perturbations}
h = 0 \eqend{,} \qquad h^{0n} = 0 \eqend{,} \qquad \partial_m h^{mn} = 0 \eqend{.}
\end{equation}

Inserting the background scale factor solution~\eqref{scale_factor} into the gravitational action~\eqref{action_grav_field_in_a_hmn}, the linear term $S^{(1)}_\text{G}$ vanishes as expected. Integrating by parts and using
\begin{equation}
\int h^{mn} f(\eta) h'_{mn} \total^4 x = - \frac{1}{2} \int h^{mn} f'(\eta) h_{mn} \total^4 x \eqend{,}
\end{equation}
the quadratic term of the gravitational action becomes
\begin{equation}
\label{grav_action_pert_in_w}
\begin{split}
4 S^{(2)}_\text{G}[a, h_{ab}] &= - \left( 1 + (5\alpha-2\beta) \kappa^2 H^2 \right) \int a^2 ( \partial^s h^{mn} ) ( \partial_s h_{mn} ) \total^4 x - 6 \alpha \kappa^2 H^2 \int a^2 h^{\prime mn} h'_{mn} \total^4 x \\
&\quad+ 3 \alpha \kappa^2 \int ( \dalembert h^{mn} ) ( \dalembert h_{mn} ) \ln a \total^4 x \eqend{.}
\end{split}
\end{equation}
The effective action $S_\text{eff}$~\eqref{general_action_in_a_hmn} can then be written as
\begin{equation}
\label{general_action_in_a_hmn_tensor}
\begin{split}
S_\text{eff}[\tilde{g}^\pm_{ab}] &= S^{(2)}_\text{G}[h^+_{ab}] - S^{(2)}_\text{G}[h^-_{ab}] + \frac{3}{2} \alpha \kappa^2 \iint \left( \dalembert h^+_{pq}(x) \right) \left( \dalembert' h^{-pq}(x') \right) K(x-x') \total^4 x \total^4 x' \\
&\quad+ \frac{3}{4} \alpha \kappa^2 \iint \left( \dalembert h^+_{pq}(x) \right) \left( \dalembert' h^{+pq}(x') \right) K^+(x-x'; \bar{\mu}) \total^4 x \total^4 x' \\
&\quad- \frac{3}{4} \alpha \kappa^2 \iint \left( \dalembert h^-_{pq}(x) \right) \left( \dalembert' h^{-pq}(x') \right) K^-(x-x'; \bar{\mu}) \total^4 x \total^4 x' \eqend{.}
\end{split}
\end{equation}
Note that we can alternatively shift the last term in the gravitational action~\eqref{grav_action_pert_in_w} into the kernels $K^\pm$, which amounts to replacing $\bar{\mu}$ with $\bar{\mu} a = - \bar{\mu}/(H \eta)$. Equation~\eqref{general_action_in_a_hmn_tensor} corresponds to equation~\eqref{effective_action_expansion_kappa} and is the starting point for the next step: computing the metric correlations.

It should be pointed out that when integrating out the matter fields, one naturally gets an effective action where, in contrast with equations~\eqref{effective_action_in_a_g} and \eqref{general_action_in_a_hmn_tensor}, the differential operators are acting on the kernels rather than the metric perturbations \cite{martin99}. If one assumes that the metric perturbations in the effective action are regular test functions, one can integrate by parts and recover those equations. This assumption is actually too restrictive when performing a calculation like ours and one gets boundary term contributions at the initial times, $t_0^+$ and $t_0^-$, from the integration by parts of the time variables (the boundary terms at the final time $T$ cancel out provided that it is larger than any other relevant times). Nevertheless, when applying the prescription described in section~\ref{overview_inin} and taking the limit $t_0^\pm \to -\infty (1 \mp \mathi \epsilon)$, such boundary terms vanish and do not contribute to our results.

\section{The gravitational two-point function}
\label{twopoint_func}

\subsection{Tree level}
\label{twopoint_treelevel}

To calculate the two-point function to first order in $V$ (which is proportional to $\kappa^2$) as given by equation~\eqref{propagator_first_order_integral}, we need to compute the lowest order propagator. For this we note that the gravitational action \eqref{grav_action_pert_in_w} is given at lowest order by
\begin{equation}
\label{lowest_order_effective_action}
S^{(2)}_\text{G}(\kappa = 0) = S_0 = - \frac{1}{4} \int a^2 ( \partial^s h^{mn} ) ( \partial_s h_{mn} ) \total^4 x \eqend{.}
\end{equation}
We now follow Ford and Parker~\cite{fordparker77} to express this action as that of two scalar fields corresponding to the two physical polarizations of the tensor perturbations. We first Fourier transform the field with respect to the spatial coordinates,
\begin{equation}
h_{mn}(x) = \int \tilde{h}_{mn}(\eta, \vec{p}) \mathe^{\mathi \vec{p} \vec{x}} \frac{\total^3 p}{(2\mathpi)^3}
\end{equation}
and note that the gauge-fixing conditions~\eqref{tensor_gauge_perturbations} for tensor perturbations correspond to
\begin{equation}
\label{tensor_gauge_perturbations_fourier}
\tilde{h} = 0 \eqend{,} \qquad \tilde{h}^{0n} = 0 \eqend{,} \qquad p_m \tilde{h}^{mn} = 0 \eqend{.}
\end{equation}
Next, we choose a right-handed orthogonal set of three (real) vectors $\vec{e}_A(\vec{p})$ such that $\vec{p} =\nolinebreak[4]\abs{\vec{p}} \vec{e}_3(\vec{p})$, and define the following two (real) scalars corresponding to the two polarizations:
\begin{equation}
\label{scalars_from_graviton}
\begin{split}
\tilde{h}_+(\eta, \vec{p}) &= \frac{\mathi}{2} \tilde{h}_{mn}(\eta, \vec{p}) \left[ e^m_1(\vec{p}) e^n_1(\vec{p}) - e^m_2(\vec{p}) e^n_2(\vec{p}) \right] \\
\tilde{h}_\times(\eta, \vec{p}) &= \tilde{h}_{mn}(\eta, \vec{p}) e^m_1(\vec{p}) e^n_2(\vec{p}) \eqend{.}
\end{split}
\end{equation}
As is well known, such a definition is not possible in a way that would be continous for every orientation of $\vec{p}$. However, with respect to the transformation $\vec{p} \to -\vec{p}$ one can adopt consistently the definition
\begin{equation}
\vec{e}_3(-\vec{p}) = - \vec{e}_3(\vec{p}) \eqend{,}\qquad \vec{e}_1(-\vec{p}) = \vec{e}_2(\vec{p}) \eqend{,}\qquad \vec{e}_2(-\vec{p}) = \vec{e}_1(\vec{p}) \eqend{.}
\end{equation}
Different definitions would be possible, but lead to the same result for the graviton propagator~\eqref{graviton_propagator}. The relation~\eqref{scalars_from_graviton} can now be inverted to give
\begin{equation}
\label{h_mn_in_h_plus_cross}
\tilde{h}^{mn}(\eta, \vec{p}) = - \mathi \tilde{h}_+(\eta, \vec{p}) \left[ e^m_1(\vec{p}) e^n_1(\vec{p}) - e^m_2(\vec{p}) e^n_2(\vec{p}) \right] + \tilde{h}_\times(\eta, \vec{p}) \left[ e^m_1(\vec{p}) e^n_2(\vec{p}) + e^m_2(\vec{p}) e^n_1(\vec{p}) \right] \eqend{.}
\end{equation}
Using the reality of $h_{mn}$, i.e. $\tilde{h}_{mn}(\eta, -\vec{p}) = \tilde{h}^*_{mn}(\eta, \vec{p})$, it follows that
\begin{equation}
\begin{split}
S_0 &= \frac{1}{2} \iint a^2(\eta) \Big[ \tilde{h}'_+(\eta, \vec{p}) \tilde{h}^{*\prime}_+(\eta, \vec{p}) - \vec{p}^2 \tilde{h}_+(\eta, \vec{p}) \tilde{h}^*_+(\eta, \vec{p}) \\
&\qquad\qquad\qquad+ \tilde{h}'_\times(\eta, \vec{p}) \tilde{h}^{*\prime}_\times(\eta, \vec{p}) - \vec{p}^2 \tilde{h}_\times(\eta, \vec{p}) \tilde{h}^*_\times(\eta, \vec{p}) \Big] \frac{\total^3 p}{(2\mathpi)^3} \total \eta \eqend{,}
\end{split}
\end{equation}
which after inverting the Fourier transform reads
\begin{equation}
\label{action_scalar_fields}
S_0 = - \frac{1}{2} \int a^2 \left[ \left( \partial^s h_+ \right) \left( \partial_s h_+ \right) + \left( \partial^s h_\times \right) \left( \partial_s h_\times \right) \right] \total^4 x \eqend{,}
\end{equation}
which is the action for a pair of minimally coupled scalar fields in a spatially flat FLRW background, whose canonically conjugate momenta are given by
\begin{equation}
\pi_+ = a^2 h'_+ \eqend{,}\qquad \pi_\times = a^2 h'_\times \eqend{.}
\end{equation}
We quantize these fields by imposing that the field operators satisfy the canonical commutation relations at equal times,
\begin{equation}
\label{canonical_commutation_plus_cross}
[ \op{h}_+(\eta, \vec{x}), \op{\pi}_+(\eta, \vec{y}) ] = [ \op{h}_\times(\eta, \vec{x}), \op{\pi}_\times(\eta, \vec{y}) ] = \mathi \delta^3(\vec{x}-\vec{y}) \eqend{,}
\end{equation}
with all other commutators vanishing. Fourier transforming the spatial coordinates, this is equivalent to
\begin{equation}
\label{canonical_commutation_plus_cross_fourier}
[ \tilde{\op{h}}_+(\eta, \vec{p}), \tilde{\op{\pi}}_+(\eta, \vec{q}) ] = [ \tilde{\op{h}}_\times(\eta, \vec{p}), \tilde{\op{\pi}}_\times(\eta, \vec{q}) ] = \mathi (2\mathpi)^3 \delta^3(\vec{p} + \vec{q}) \eqend{.}
\end{equation}
From the action~\eqref{action_scalar_fields} it follows that the equations of motion for the operators in the Heisenberg picture are
\begin{equation}
\label{h_plus_cross_eom}
\Big( \dalembert - 2 H a \partial_\eta \Big) \op{h}_+ = \Big( \dalembert - 2 H a \partial_\eta \Big) \op{h}_\times = 0 \eqend{.}
\end{equation}

To calculate the propagator, we need to find a complete set of solutions (``mode functions'') to those equations, which then determine the expansion of the perturbation operators $\op{h}_+$ and $\op{h}_\times$ and the associated vacuum state. In the following, we will perform the analysis for $\op{h}_+$ and trivially extend the results for $\op{h}_\times$. Fourier transforming the first equation in~\eqref{h_plus_cross_eom} with respect to spatial coordinates, it becomes
\begin{equation}
\label{equation_order_zero_f}
\tilde{\op{h}}''_+(\eta, \vec{p}) - \frac{2}{\eta} \tilde{\op{h}}'_+(\eta, \vec{p}) + \vec{p}^2 \tilde{\op{h}}_+(\eta, \vec{p}) = 0 \eqend{,}
\end{equation}
whose general solution is given by
\begin{equation}
\label{h_plus_expansion_creation_annihilation}
\tilde{\op{h}}_+(\eta, \vec{p}) = C(\vec{p}) \hat{a}_+(\vec{p}) ( \abs{\vec{p}} \eta - \mathi ) \mathe^{- \mathi \abs{\vec{p}} \eta} + C^*(-\vec{p}) \hat{a}^\dagger_+(-\vec{p}) ( \abs{\vec{p}} \eta + \mathi ) \mathe^{\mathi \abs{\vec{p}} \eta}
\end{equation}
with an arbitrary operator $\hat{a}_+$ and a normalization constant $C$, and where we have used the reality condition for $h_+(x)$, which corresponds to hermiticity for $\op{h}_+(x)$. For the canonical momentum, we get
\begin{equation}
\tilde{\op{\pi}}_+(\eta, \vec{p}) = - \mathi \frac{\vec{p}^2}{H^2 \eta} \left[ C(\vec{p}) \hat{a}_+(\vec{p}) \mathe^{- \mathi \abs{\vec{p}} \eta} - C^*(-\vec{p}) \hat{a}^\dagger_+(-\vec{p}) \mathe^{\mathi \abs{\vec{p}} \eta} \right] \eqend{.}
\end{equation}

The canonical commutation relations~\eqref{canonical_commutation_plus_cross_fourier}, together with the vanishing of the equal-time commutators for $\op{h}_+$ and $\op{\pi}_+$ with themselves, are valid if and only if
\begin{equation}
\begin{split}
C(\vec{p}) C^*(\vec{p}) &= C(-\vec{p}) C^*(-\vec{p}) \\
[ \hat{a}_+(\vec{p}) , \hat{a}_+(\vec{q}) ] &= [ \hat{a}^\dagger_+(\vec{p}) ,  \hat{a}^\dagger_+(\vec{q}) ] = 0 \\
[ \hat{a}_+(\vec{p}), \hat{a}^\dagger_+(\vec{q}) ] &= \frac{(2\mathpi)^3 \delta^3(\vec{p} - \vec{q}) H^2}{2 \abs{\vec{p}}^3 C(\vec{p}) C^*(\vec{p})} \eqend{.}
\end{split}
\end{equation}
By choosing $C(\vec{p}) = H \left( 2 \abs{\vec{p}}^3 \right)^{-\frac{1}{2}}$, the operators $\hat{a}_+$ and $\hat{a}^\dagger_+$ fulfill the standard commutation relations for creation and annihilation operators. We then define the free field vacuum $\ket{0}$ by
\begin{equation}
\hat{a}_+(\vec{p}) \ket{0} = \hat{a}_\times(\vec{p}) \ket{0} = 0 \eqend{.}
\end{equation}
This vacuum which (for $\abs{\vec{p}} \eta \gg 1$) resembles the Minkowski vacuum with respect to the choice of mode functions~\eqref{h_plus_expansion_creation_annihilation}, is the Bunch-Davies vacuum~\cite{bunchdavies78} for the gravitons, and we will use it in the following (there is a subtelty in this case concerning IR divergences which is briefly discussed below).

The lowest order two-point function of the tensor perturbations $\bra{0} \op{h}^{ab}(x) \op{h}^{cd}(y) \ket{0}$ can now be evaluated directly using the decomposition~\eqref{h_mn_in_h_plus_cross} into the $+$ and $\times$ polarizations and the mode expansion~\eqref{h_plus_expansion_creation_annihilation}. A straightforward calculation then leads to
\begin{equation}
\label{graviton_propagator}
\bra{0} \tilde{\op{h}}^{ab}(\eta, \vec{p}) \tilde{\op{h}}^{cd}(\eta', \vec{q}) \ket{0} = (2\mathpi)^3 \delta^3(\vec{p} + \vec{q}) f(\eta, \eta', \abs{\vec{p}}) P^{abcd}(\vec{p}) \eqend{,}
\end{equation}
where
\begin{equation}
f(\eta, \eta', \abs{\vec{p}}) = \frac{H^2}{2 \abs{\vec{p}}^3} ( \abs{\vec{p}} \eta - \mathi ) ( \abs{\vec{p}} \eta' + \mathi ) \mathe^{- \mathi \abs{\vec{p}} (\eta-\eta')} = f^*(\eta', \eta, \abs{\vec{p}}) \eqend{,}
\end{equation}
and
\begin{equation}
\label{TTprojector}
\begin{split}
P^{abcd}(\vec{p}) &= \left( e^a_1(\vec{p}) e^b_2(\vec{p}) + e^a_2(\vec{p}) e^b_1(\vec{p}) \right) \left( e^c_1(\vec{p}) e^d_2(\vec{p}) + e^c_2(\vec{p}) e^d_1(\vec{p}) \right) \\
&\quad+ \left( e^a_1(\vec{p}) e^b_1(\vec{p}) - e^a_2(\vec{p}) e^b_2(\vec{p}) \right) \left( e^c_1(\vec{p}) e^d_1(\vec{p}) - e^c_2(\vec{p}) e^d_2(\vec{p}) \right) \\
&= P^{ad} P^{bc} + P^{ac} P^{bd} - P^{ab} P^{cd}
\end{split}
\end{equation}
is the polarization tensor which can be written in terms of the projection tensor $P^{ab}$ defined by
\begin{equation}
P^{ab} = \eta^{ab} + \delta_0^a \delta_0^b - \frac{p^a p^b}{\vec{p}^2} \eqend{,}
\end{equation}
where $p^a = (0, \vec{p})$ in the coordinate system that we have chosen. The projection tensor $P^{ab}$ satisfies
\begin{equation}
P^{ab} \eta_{ab} = 2 \eqend{,}\qquad P^{ab} P^{cd} \eta_{bc} = P^{ad} \eqend{.}
\end{equation}

From equation~\eqref{graviton_propagator} we can now calculate all the lowest order CTP propagators for the graviton. They are given by
\begin{equation}
\tilde{G}^{0AB}_{abcd}(\eta, \eta', \vec{p}) = \tilde{G}^{0AB}(\eta, \eta', \vec{p}) P_{abcd}(\vec{p}) \eqend{,}
\end{equation}
where the components of the $2\times2$ matrix $\tilde{G}^0$ are
\begin{equation}
\label{propagators_at_order_zero}
\begin{split}
\tilde{G}^{0++}(\eta, \eta', \vec{p}) &= - \mathi \left[ \Theta(\eta - \eta') f(\eta, \eta', \abs{\vec{p}}) + \Theta(\eta' - \eta) f^*(\eta, \eta', \abs{\vec{p}}) \right] \\
\tilde{G}^{0+-}(\eta, \eta', \vec{p}) &= - \mathi f^*(\eta, \eta', \abs{\vec{p}}) \\
\tilde{G}^{0-+}(\eta, \eta', \vec{p}) &= - \mathi f(\eta, \eta', \abs{\vec{p}}) \\
\tilde{G}^{0--}(\eta, \eta', \vec{p}) &= - \mathi \left[ \Theta(\eta - \eta') f^*(\eta, \eta', \abs{\vec{p}}) + \Theta(\eta' - \eta) f(\eta, \eta', \abs{\vec{p}}) \right] \eqend{.}
\end{split}
\end{equation}

Since $f$ goes like $\abs{\vec{p}}^{-3}$ for small $\abs{\vec{p}}$, this two-point function must be truncated at some small momentum $\abs{\vec{p}} = \xi$ to have a well-defined inverse Fourier transform. The two-point function in coordinate space diverges then as $\ln \xi$ for $\xi \to 0$, as was noted by a number of authors~\cite{gravitonprop1,gravitonprop2}. This was to be expected since the same problem occurs for a massless minimally coupled scalar field in de Sitter space~\cite{massless1,massless2,massless3}. However, similarly to the stress tensor correlation function for a scalar field \cite{massless3}, gauge-invariant correlation functions of observables characterizing the local geometry in a finite-size region, such as the curvature tensor correlation functions, are finite and do not grow at large distances~\cite{massless3,weylcorrelation,curvaturecorr}. We will not discuss the issue in this paper, but plan to address it in a future publication~\cite{frv2012a}.

\subsection{One loop}
\label{twopoint_oneloop}

We can now use equation~\eqref{propagator_first_order_integral} to calculate the one-loop order graviton propagators which are of order $\kappa^2$ in our perturbative scheme. As explained at the end of section~\ref{overview_calculus}, this is easily accomplished by replacing the fields $h_{ab}$ in the effective action \eqref{general_action_in_a_hmn_tensor} with the lowest order propagators. Fourier transforming the spatial coordinates, we get
\begin{equation}
\label{propagator_at_first_order}
\tilde{G}^{AB}_{abcd}(\eta, \eta', \vec{p}) = \tilde{G}^{0AB}_{abcd}(\eta, \eta', \vec{p}) + P_{abcd}(\vec{p}) \kappa^2 \tilde{G}^{1AB}(\eta, \eta', \vec{p}) \\
\end{equation}
where
\begin{equation}
\label{propagator_at_first_order_two}
\begin{split}
\tilde{G}^{1AB}(\eta, \eta', \vec{p}) &= (5\alpha-2\beta) \vec{p}^2 \int \tau^{-2} \tilde{G}^0_{AM}(\eta, \tau, \vec{p}) \epsilon_{MN} \tilde{G}^0_{NB}(\tau, \eta', \vec{p}) \total \tau \\
&\quad+ (\alpha+2\beta) \int \tau^{-2} ( \partial_\tau \tilde{G}^0_{AM}(\eta, \tau, \vec{p}) ) \epsilon_{MN} ( \partial_\tau \tilde{G}^0_{NB}(\tau, \eta', \vec{p}) ) \total \tau \\
&\quad+ 3 \alpha \int \left[ ( \partial^2_\tau + \vec{p}^2 ) \tilde{G}^0_{AM}(\eta, \tau, \vec{p}) \right] \epsilon_{MN} \left[ ( \partial^2_\tau + \vec{p}^2 ) \tilde{G}^0_{NB}(\tau, \eta', \vec{p}) \right] \ln (- H \tau) \total \tau \\
&\!\!\!\!\!\!\!\!- \frac{3}{2} \alpha \iint \left[ ( \partial^2_\tau + \vec{p}^2 ) \tilde{G}^0_{AM}(\eta, \tau, \vec{p}) \right] V_{MN}(\tau, \tau', \vec{p}) \left[ ( \partial^2_{\tau'} + \vec{p}^2 ) \tilde{G}^0_{NB}(\tau', \eta', \vec{p}) \right] \total \tau \total \tau' \eqend{,}
\end{split}
\end{equation}
$\epsilon_{MN} = \begin{pmatrix} 1 & 0 \\ 0 & -1 \end{pmatrix}$ and
\begin{equation}
V_{MN}(\tau, \tau', \vec{p}) = \begin{pmatrix} \tilde{L}\left(\tau-\tau', \vec{p}; \bar{\mu} \right) & \tilde{D}(\tau-\tau', \vec{p}) \\ - \tilde{D}(\tau-\tau', \vec{p}) & - \tilde{L}\left(\tau-\tau', \vec{p}; \bar{\mu} \right) \end{pmatrix} + \mathi \tilde{N}(\tau-\tau', \vec{p}) \begin{pmatrix} 1 & - 1 \\ - 1 & 1 \end{pmatrix} \eqend{.}
\end{equation}
This is written in terms of the Fourier transforms of the following real-valued kernels, in terms of which the complex-valued kernels in~\eqref{kernels_def} can be decomposed:
\begin{equation}
\label{def_kernels_zwei}
\begin{split}
L(x; \bar{\mu}) &= - \int \ln \abs{\frac{p^2}{\bar{\mu}^2}} \mathe^{\mathi p x} \frac{\total^4 p}{(2\mathpi)^4} = L(-x; \bar{\mu}) \\
N(x) &= \mathpi \int \Theta(-p^2) \mathe^{\mathi p x} \frac{\total^4 p}{(2\mathpi)^4} = N(-x) \\
D(x) &= \mathi \mathpi \int \Theta(-p^2) \sgn p^0 \mathe^{\mathi p x} \frac{\total^4 p}{(2\mathpi)^4} = -D(-x) \eqend{.}
\end{split}
\end{equation}
The Fourier transforms of those kernels can be calculated with the help of appendix~\ref{appendix_distributions} and are given by
\begin{equation}
\label{kernels_fts}
\begin{split}
\tilde{L}(\eta-\eta', \vec{p}) &= \cos \left[ \abs{\vec{p}} (\eta-\eta') \right] \mathcal{P}' \frac{1}{\abs{\eta-\eta'}} \\
\tilde{D}(\eta-\eta', \vec{p}) &= \cos \left[ \abs{\vec{p}} (\eta-\eta') \right] \mathcal{P} \frac{1}{\eta-\eta'} \\
\tilde{N}(\eta-\eta', \vec{p}) &= - \frac{\sin \left[ \abs{\vec{p}} (\eta-\eta') \right]}{\eta - \eta'} + \mathpi \delta(\eta-\eta') \eqend{.}
\end{split}
\end{equation}

After calculating the integrals using the $\mathi \epsilon$ prescription as explained in some detail in appendix~\ref{appendix_integrals}, we get for the two-point function of the metric perturbations
\begin{equation}
\label{two_point_function_h}
\begin{split}
\bra{\text{in}} \tilde{\op{h}}^{ab}(\eta, \vec{p}) \tilde{\op{h}}^{mn}(\eta', \vec{q}) \ket{\text{in}} &= (2\mathpi)^3 \delta^3(\vec{p}+\vec{q}) P^{abmn} \times \\
&\times \bigg[ f(\eta, \eta', \vec{p}) \left( 1 + 6 \alpha \kappa^2 H^2 \ln\left( \frac{\bar{\mu}}{H} \right) - \left( 5 \alpha - 2 \beta \right) \kappa^2 H^2 \right) \\
&\qquad+ \frac{3}{2} \alpha \kappa^2 H^4 \Big( I_1(\eta, \eta', \vec{p}) - I_2(\eta, \eta', \vec{p}) - I^*_2(\eta', \eta, \vec{p}) \Big) \\
&\qquad+ \frac{3}{2} \alpha \kappa^2 H^4 \Big( I_3(\eta, \eta', \vec{p}) - I_4(\eta, \eta', \vec{p}) + I_5(\eta, \eta', \vec{p}) \Big) \bigg] + \bigo{\kappa^4} \eqend{,}
\end{split}
\end{equation}
where
\begin{equation}
\begin{split}
I_1(\eta, \eta', \vec{p}) &= 2 \abs{\vec{p}}^{-1} \eta \eta' \mathe^{- \mathi \abs{\vec{p}} (\eta-\eta')} \\
I_2(\eta, \eta', \vec{p}) &= \abs{\vec{p}}^{-3} \mathe^{\mathi \abs{\vec{p}} (\eta+\eta')} \left( \abs{\vec{p}} \eta + \mathi \right) \left( \abs{\vec{p}} \eta' + \mathi \right) \left[ \Ein\left( - 2 \mathi \abs{\vec{p}} \eta \right) + \ln\left( 2 \mathi \abs{\vec{p}} \eta \right) + \gamma \right] \\
&\qquad+ \abs{\vec{p}}^{-3} \mathe^{- \mathi \abs{\vec{p}} (\eta-\eta')} \left( \abs{\vec{p}} \eta - \mathi \right) \left( \abs{\vec{p}} \eta' + \mathi \right) \ln\left( - 2 \abs{\vec{p}} \eta \right) \\
I_3(\eta, \eta', \vec{p}) &= \abs{\vec{p}}^{-3} \mathe^{- \mathi \abs{\vec{p}} (\eta-\eta')} \left( \abs{\vec{p}} \eta - \mathi \right) \left( \abs{\vec{p}} \eta' + \mathi \right) \left[ \ln\left( 2 \mathi \abs{\vec{p}} (\eta - \eta') \right) + \gamma \right] \\
I_4(\eta, \eta', \vec{p}) &= \abs{\vec{p}}^{-3} \mathe^{\mathi \abs{\vec{p}} (\eta-\eta')} \left( \abs{\vec{p}} \eta + \mathi \right) \left( \abs{\vec{p}} \eta' - \mathi \right) \left[ \Ein\left( - 2 \mathi \abs{\vec{p}} (\eta-\eta') \right) + \ln\left( 2 \mathi \abs{\vec{p}} (\eta - \eta') \right) + \gamma \right] \\
I_5(\eta, \eta', \vec{p}) &= \eta^2 (\eta')^2 \left[ \tilde{N}(\eta-\eta', \vec{p}) - \mathi \tilde{D}(\eta-\eta', \vec{p}) \right] \eqend{.}
\end{split}
\end{equation}
This is the exact result for the one-loop-corrected (due to conformal matter fields) two-point function for tensor metric perturbations, and unlike previous results, it has been obtained for arbitrary pairs of times rather than just equal times. It is finite for non-coincident points and (apart from a global factor $\abs{\vec{p}}^{-3}$) invariant under the simultaneous rescaling $\{ \vec{p} \to \lambda^{-1} \vec{p}; \eta, \eta' \to \lambda \eta, \lambda \eta' \}$, which is a necessary (but not sufficient) condition for de Sitter invariance.

The origin of the two kernels $\tilde{N}$ and $\tilde{D}$ in the result for $I_5(\eta, \eta', \vec{p})$ is the following. The two CTP propagators $G_{AB}$ in equation~\eqref{propagator_at_first_order_two}, or equivalently equation~\eqref{j_integral}, appear with a linear operator acting on them which contains second-order time derivatives. If $G_{AB}$ is a true propagator rather than a Wightman function, i.e., for $G_{++}$ and $G_{--}$ (or $G_\text{ret}$ in appendix~\ref{appendix_integrals}), the action of the linear operator gives rise to a term proportional to a Dirac $\delta$ distribution in time. When that is the case for both $G_{AB}$ and $G_{CD}$ in equations~\eqref{propagator_at_first_order_two} or~\eqref{j_integral}, the result of the two time integrals is simply a term proportional to the corresponding kernel, evaluated at the two times $\eta$ and $\eta'$ in the argument of the metric correlation function.

Equation~\eqref{two_point_function_h} is valid for the particular choice of the renormalization scale such that $a_1^\text{ren}(\bar{\mu}) = 0$, as explained in section~\ref{effective_action}. For arbitrary values of the renormalization scale one should make the following replacement in equation~\eqref{two_point_function_h}:
\begin{equation}
\label{replacement}
6 \alpha \kappa^2 H^2 \ln\left( \frac{\bar{\mu}}{H} \right) \  \longrightarrow \  6 \kappa^2 H^2 \left[ N a_1^\text{ren}(\mu) + \alpha \ln\left( \frac{\mu}{H} \right) \right] \eqend{.}
\end{equation}
Note that this guarantees the invariance of the two-point function under the renormalization group, as follows from equation~\eqref{renormalized_parameter}.

\subsection{The power spectrum}
\label{twopoint_spectrum}

In cosmology, two-point correlations are often expressed in terms of the power spectrum. To calculate it, we need to take the equal time limit $\eta' \to \eta$ of the expectation value~\eqref{two_point_function_h}. However, this expectation value is not a regular function but a distribution with singular support, for which the equal-time limit does not make sense. Therefore, the concept of a power spectrum is, strictly speaking, ill-defined at the one-loop level (see appendix~\ref{appendix_coincidence} for a more detailed discussion of this point). In order to compare with previous results in the literature, however, we can take the equal time limit of the regular part, which amounts to replacing the kernels appearing in $I_5$ by
\begin{equation}
\begin{split}
\tilde{N}(\eta, \eta', \vec{p}) &\to - \frac{\sin \left[ \abs{\vec{p}} (\eta-\eta') \right]}{\eta - \eta'} \\
\tilde{D}(\eta-\eta', \vec{p}) &\to 0 \eqend{.}
\end{split}
\end{equation}
Following section 18 in reference~\cite{perturbations} (and taking into account that our metric perturbations and theirs are related by a factor $\kappa$), the power spectrum is defined to be a suitable normalization factor times the spatial Fourier transform of the contracted correlation function at equal times:
\begin{equation}
\label{powerspectrum_def}
\begin{split}
\delta^2(\abs{\vec{p}}, \eta) &= \frac{\kappa^2}{4 (2\mathpi)^3} \abs{\vec{p}}^3 \eta_{am} \eta_{bn} \int \bra{\text{in}} h^{ab}(\eta, \vec{x}) h^{mn}(\eta, \mathbf{0}) \ket{\text{in}} \mathe^{- \mathi \vec{p} \vec{x}} \total^3 x \\
&= \frac{\kappa^2}{32 \mathpi^3} \abs{\vec{p}}^3 \eta_{am} \eta_{bn} \int \bra{\text{in}} \tilde{h}^{ab}(\eta, \vec{p}) \tilde{h}^{mn}(\eta, \vec{q}) \ket{\text{in}} \frac{\total^3 q}{(2\mathpi)^3} \eqend{,}
\end{split}
\end{equation}
which gives
\begin{equation}
\label{power_spectrum}
\begin{split}
\delta^2(\abs{\vec{p}}, \eta) &= \frac{\kappa^2 H^2}{32 \mathpi^3} \bigg[ 2 \left( 1 + \vec{p}^2 \eta^2 \right) \left( 1 + 6 \alpha \kappa^2 H^2 \ln\left( \frac{\bar{\mu}}{H} \right) - \left( 5 \alpha - 2 \beta \right) \kappa^2 H^2 \right) \\
&\qquad- 12 \alpha \kappa^2 H^2 \Re \left[ \mathe^{2 \mathi \abs{\vec{p}} \eta} \left( \abs{\vec{p}} \eta + \mathi \right)^2 \big[ \Ein\left( - 2 \mathi \abs{\vec{p}} \eta \right) + \ln\left( 2 \mathi \abs{\vec{p}} \eta \right) + \gamma \big] \right] \\
&\qquad- 12 \alpha \kappa^2 H^2 \left( 1 + \vec{p}^2 \eta^2 \right) \ln\left( - 2 \abs{\vec{p}} \eta \right) + 6 \alpha \kappa^2 H^2 \vec{p}^2 \eta^2 \left( 2 - \vec{p}^2 \eta^2 \right) \bigg] + \bigo{\kappa^6} \eqend{.}
\end{split}
\end{equation}
Once again, for an arbitrary value of the renormalization scale $\mu$ one needs the replacement~\eqref{replacement}.

Let us consider two important limits of equation~\eqref{power_spectrum}. It is useful to write it in terms of the physical momentum (of the physical spacetime $\tilde{g}_{ab}$), which is
\begin{equation}
\tilde{p} = \frac{\abs{\vec{p}}}{a} = - H \abs{\vec{p}} \eta \eqend{.}
\end{equation}
The power spectrum depends on $\abs{\vec{p}}$ and $\eta$ only through the combination $\tilde{p}$, so that it is time-independent in the physical spacetime.

For ``sub-horizon'' modes with $\tilde{p} \gg H$ we should then recover flat space behavior, as the curvature of spacetime should have no effect on those modes. Taking therefore the limit $H \to 0$, we get from equation~\eqref{power_spectrum}
\begin{equation}
\delta^2(\tilde{p}) \to \frac{\kappa^2}{16 \mathpi^3} \tilde{p}^2 \left( 1 - 3 \alpha \kappa^2 \tilde{p}^2 + \bigo{\kappa^4} \right) + \bigo{\frac{H}{\tilde{p}}} \eqend{,}
\end{equation}
where we have used the asymptotic expansion of the $\Ein$ function from appendix~\ref{appendix_special}. We therefore have a power correction to the Minkowski spectrum \cite{durrer09}.

On the other hand, for ``super-horizon'' modes with $\tilde{p} \ll H$ we take the limit $\tilde{p} \to 0$ to find
\begin{equation}
\label{super_horizon}
\delta^2(\tilde{p}) \to \frac{\kappa^2 H^2}{16 \mathpi^3} \left[ 1 + 6 \alpha \kappa^2 H^2 \left( \ln\left( \frac{\bar{\mu}}{H} \right) + \gamma \right) - (5\alpha - 2\beta) \kappa^2 H^2 + \bigo{\kappa^4} \right] + \bigo{\frac{\tilde{p}}{H}} \eqend{,}
\end{equation}
which corresponds to a small constant shift of the scale-invariant power spectrum for tensor perturbations in de Sitter obtained at tree level (for arbitrary renormalization scale $\mu$, one needs again the replacement \eqref{replacement}). This is in contrast with the logarithmic dependence on the comoving momentum $\abs{\vec{p}}$ previously found for one-loop corrections to tensor \cite{aelim09} and scalar \cite{weinberg05,chaicherdsakul07} perturbations, but is in line with the more recent results on loop corrections for scalar perturbations due to minimally coupled fields obtained in \cite{senatore}, where the origin of this discrepancy was identified.

\subsection{Cancellation of secular terms}
\label{secular_terms}

In our result for the one-loop two-point function of the tensor metric perturbations, there is no dependence left on the initial times $t_0^\pm = t_0 (1 \mp \mathi \epsilon)$ after taking the limit $t_0 \to -\infty$. Being able to eliminate any dependence on $t_0$ is crucial for obtaining a result which is invariant (aside from a global factor $\abs{\vec{p}}^{-3}$) under the simultaneous rescaling $\{\vec{p} \to \lambda^{-1} \vec{p};\, \eta,\eta' \to \lambda\eta, \lambda\eta'\}$, which is a necessary condition for the de Sitter invariance of our result. In contrast, a related study of one-loop corrections to tensor perturbations around de Sitter \cite{wuetal11} found a result which is particularly sensitive to large values of $t_0$ and diverges in the limit $t_0 \to -\infty$. We provide here a simple explanation of the origin of those terms and why they are absent in our case.

It is useful to consider the following decomposition for the anticommutator of the metric perturbations when the effects of matter loops are included:
\begin{equation}
\label{stochastic}
\begin{split}
\frac{1}{2} \expect{ \big\{ \hat{h}_{ab}(x), \hat{h}_{cd}(x') \big\} } &= \expect{ h_{ab}^\text{h}(x) h_{cd}^\text{h}(x') }_0 \\
&\qquad+ \frac{\kappa^2}{4} \iint G^\mathrm{ret}_{abmn}(x, y) N^{mnpq}(y,y') G^\mathrm{ret}_{pqcd}(y', x') \sqrt{-g} \total^4 y' \sqrt{-g} \total^4 y \eqend{,}
\end{split}
\end{equation}
where the noise kernel $N_{abcd}(x,y)$ corresponds to the symmetrized connected correlation function of the stress tensor
\begin{equation}
N_{abcd}(x,x') = \frac{1}{2} \bra{0} \{ \hat{T}_{ab}(x), \hat{T}_{cd}(x') \} \ket{0} - \bra{0} \hat{T}_{ab}(x) \ket{0} \bra{0} \hat{T}_{cd}(x') \ket{0} \eqend{.}
\end{equation}
In addition, $h_{ab}^\text{h}(x)$ correspond to solutions of the linearized semiclassical Einstein equations $G_{ab}^{(1)}\left[g+h\right] = \frac{1}{2} \kappa^2 \expect{ \hat{T}_{ab}^{(1)} [g+h] }_\text{ren}$ around a background metric $g_{ab}$, $\expect{\ldots}_0$ denotes an average over their initial conditions according to the initial quantum state of the metric perturbations, and $G^\text{ret}_{abcd}$ is the retarded propagator associated with the linearized semiclassical equation.
Equation~\eqref{stochastic} can be derived using the stochastic gravity formalism, but also within a purely quantum field theoretical calculation \cite{hrv_orderredux,martin99,hrv04,stochasticgravity} and assumes a factorized initial state with the metric perturbations and the matter fields uncorrelated.\footnote{Such kinds of initial states give rise to certain pathologies \cite{hrv_orderredux,QBM}. Hence, in our study we have employed the method described in section~\ref{overview_inin} to generate a properly dressed initial state for the interacting theory (with correlations between the metric perturbations and the matter fields).}
The first term on the right-hand side of equation~\eqref{stochastic} can be interpreted as corresponding to the fluctuations of the metric due to fluctuations of their quantum initial state (including the effect of matter loops on their evolution as given by the linearized semiclassical equation) and are often referred to as \emph{intrinsic fluctuations}. The second term corresponds to the metric fluctuations \emph{induced} by the fluctuations of the matter fields.

It turns out that the contributions from intrinsic and induced fluctuations exhibit \emph{secular} terms when considered separately. The basic idea can be illustrated with the simple model of two bilinearly coupled harmonic oscillators \cite{crv03,QBM} with action
\begin{equation}
\label{qbm1}
S[Q,q] = \frac{1}{2} \int \left( M \dot{Q}^2 - M \Omega^2 Q^2 + m \dot{q}^2 - m \omega^2 q^2 - 2 c\, Q q \right) \total \eta
\eqend{,}
\end{equation}
where $\eta$ denotes the time and overdots denote derivatives with respect to it. Here the oscillator $q$ is the analogue of the matter fields, while the oscillator $Q$ is the analogue of the metric perturbations. The analogous quantity to consider is the correlation function $\frac{1}{2} \expect{ \{ \hat{Q}(\eta),\hat{Q}(\eta') \} }$ calculated perturbatively up to quadratic order in $c$. The free retarded propagator for the oscillator $Q$ and the noise kernel characterizing the fluctuations of the oscillator $q$ are given by
\begin{equation}
\label{qbm2}
\begin{split}
G^\mathrm{ret} (\eta-\eta') &= \frac{1}{M \Omega} \sin \left[ \Omega(\eta-\eta') \right]
\theta(\eta-\eta') \\
N(\eta-\eta') &= \frac{1}{2 m \omega} \cos\left[ \omega(\eta-\eta') \right] \eqend{,}
\end{split}
\end{equation}
where the noise kernel corresponds to the real part (the anticommutator) of the two-point function of the oscillator $q$ in the interaction picture and we have assumed that it is initially in its ground state, so that $\expect{ \{ \hat{q}(\eta),\hat{q}(\eta') \} } = (2m\omega)^{-1} \exp[ \mathi \Omega(\eta-\eta')]$. In that case the term for the induced fluctuations is proportional to
\begin{equation}
\label{induced}
\int_{t_0}^{\eta} \int_{t_0}^{\eta'} \sin\left[ \Omega(\eta-\tau) \right] \cos\left[ \omega(\tau-\tau') \right] \sin\left[ \Omega(\eta'-\tau') \right] \total \tau' \total \tau \eqend{.}
\end{equation}
The key point is that the sines and cosines involve exponentials with two opposite signs in the exponent. For resonant oscillators (with $\omega = \Omega$) this will give rise to terms for which the $\tau$- or $\tau'$-dependent exponents (or both) cancel out and one is left with terms that do not oscillate as a function of $\tau$ or $\tau'$. Therefore, when integrating over these variables, one obtains \emph{secular} terms proportional to $\eta-t_0$, $\eta'-t_0$ or both. The structure of the induced fluctuations for the loop diagram associated with the cubic gravitational interaction between the metric perturbation and the matter field is completely analogous, with a sum over the spatial momentum running in the loop, and for massless fields there is always some momentum for which the resonance condition is fulfilled.

In fact, given a spatial Fourier mode $\vec{p}$ for the metric perturbations, the expression for the one-loop correction to its two-point function is the same as the contribution quadratic in the coupling constant $c$ for a model consisting of an oscillator with $\Omega = \abs{\vec{p}}$ interacting independently (and bilinearly) with a set of oscillators $\{q_j\}$ with frequencies $\{ \omega_j \}$. The result is a simple generalization of the two-oscillator model discussed above with an extra sum over the frequencies $\omega_j$. More precisely, one needs to integrate over a continuous $\omega$ distribution, as dictated by the spatial Fourier transform of equations~\eqref{def_kernels_zwei}, which involves just an integral over $p^0$ and whose result is given by equations~\eqref{kernels_fts}. In addition, to mimic the gravitational case for metric perturbations around de Sitter, one needs to consider the coupling between the oscillators to be time-dependent with terms proportional to $\eta$. (One should also consider the frequency of the oscillator $Q$ to be time dependent, but there the time-dependent terms are proportional to $\eta^{-2}$. Therefore, they are essentially irrelevant for the discussion about secular terms because they vanish as $\eta \to -\infty$.)

Despite the existence of secular terms for the intrinsic and the induced fluctuations, they are absent in the result for the full correlation function because the secular terms from both contributions cancel out. Such cancellation can be inferred from our calculations and the final result in section~\ref{twopoint_oneloop}, but can also be easily understood when considering the Feynman-Stückelberg diagrams in the CTP formalism following the approach described in section~\ref{overview_inin}. The basic idea is that for fixed arguments of the correlation function (associated with the external legs) and integrating over the times of internal vertices, all the propagators leaving a given vertex on the ``$+$'' branch are of positive frequency\footnote{In the gravitational case this is a consequence of $G^+(\eta_1,\eta)$ being the Wightman function of the free theory associated with the Bunch-Davies vacuum, which involves an exponential factor with exponent $-\mathi \Omega (\eta_1-\eta)$. In that case there are also factors involving powers of the conformal times, but the key role is still played by the oscillatory factors as far as the possible existence of secular terms is concerned.} with respect to the time variable of that vertex at sufficiently early times (the analogous statement with negative frequencies holds for vertices on the ``$-$'' branch). Therefore, instead of an equation like \eqref{induced}, one gets a sum of terms of the following form (or analogous forms):
\begin{equation}
\label{ctp}
\int_{t_0^+}^{\eta} \int_{\tau}^{\eta'} \mathe^{-\mathi \Omega (\eta-\tau)}\, \mathe^{-\mathi \omega(\tau'-\tau)}\, \mathe^{-\mathi \Omega(\eta'-\tau')} \total \tau' \total \tau \eqend{.}
\end{equation}
where in this particular case all the $\tau$-dependent exponents have the same sign. This guarantees that there will be no secular terms and the limit $t_0 \to -\infty$ will converge. It should be emphasized that within this formulation the terms that naturally appear, such as \eqref{ctp}, always contain a mixture of intrinsic and induced contributions.
Note also that it is important to distinguish the initial times $t_0^+$ and $t_0^-$ for the two branches. Otherwise we would miss the exponential factor in terms proportional to $\mathe^{\mathi (\omega + \Omega) (t_0^+ - t_0^-)}$, or $t_0^\pm \mathe^{\mathi (\omega + \Omega) (t_0^+ - t_0^-)}$ (appearing when one has a time-dependent coupling, like the cubic interaction terms in equation~\eqref{EHaction} --- quadratic in the matter fields and linear in the metric perturbations --- for a nontrivial FLRW background). Missing these exponential factors gives rise to spurious finite contributions or even divergences in the limit $t_0 \to -\infty$ for the second case.

An alternative way of understanding the cancellation of the secular terms is by noticing that the calculation of correlation functions using the method described in section~\ref{overview_inin} gives the correlation function for the ground state of the interacting theory (or an adiabatic ground state for suitable time-dependent interactions). This is clearer for time-independent Hamiltonians (including the interactions). The ground state is then a stationary state and the time dependence of the correlation functions only involves the relative differences between their arguments. Therefore, there cannot be secular terms (depending on some particular initial time) in the exact result, which implies a complete cancellation of secular terms order by order in the perturbative calculation. The lack of dependence on an arbitrary initial time and the corresponding need for a cancellation of the secular terms can be extended to the case of time-dependent interactions provided that appropriate adiabaticity conditions are satisfied.

After the discussion in this subsection, the reason for the sensitivity to large values of $t_0$ of the result in \cite{wuetal11} is now clear: since only induced fluctuations were considered, there were secular terms analogous to those generated in equation~\eqref{induced} which did not cancel out. Note that in a similar calculation for flat space all the secular terms in the induced metric perturbations do cancel out. This is, however, an accident of such a special case:\footnote{The reason for the cancellation when considering the gravitational case in a Minkowski background \cite{martin00} is that the noise kernel for a massless field involves two global d'Alembertian operators which can be integrated by parts so that they act on the graviton propagator. The action of the Minkowski d'Alembertian on the graviton propagator gives a vanishing result in flat space, but this does not happen in a general FLRW spacetime.} in general other cubic interaction terms (different from the gravitational interaction) will not lead to such a cancellation even in flat space.
Neither is this a consequence of considering a sufficiently simple situation: there would be no cancellation for simpler non-derivative cubic couplings or even for the simplest case of two bilinearly coupled resonant harmonic oscillators with a time-independent coupling discussed above.

\section{Discussion}
\label{discussion}

In this paper we have obtained the exact result for the gravitational wave spectrum in the de Sitter spacetime including the one-loop corrections from a conformal scalar field. Our result is compatible with de Sitter invariance: besides spatial homogeneity and isotropy, it respects the isometry corresponding to a global rescaling of the 
spatial coordinates and the conformal time in spatially flat coordinates for the Poincaré patch. In particular it contains no terms involving the logarithm of the co-moving momentum, which had been found in previous calculations of the one-loop correction to the power spectrum for both scalar \cite{weinberg05,chaicherdsakul07} and tensor perturbations \cite{aelim09}. Such explicit dependence on co-moving rather than physical momentum is incompatible with the rescaling symmetry mentioned above, as pointed out in \cite{senatore}, and can lead to significant deviations from the standard picture for inflationary models lasting for a sufficiently long time. This problematic term is canceled out when the additional contribution from the spacetime volume measure is properly taken into account in dimensional regularization, which gives rise to a term proportional to the logarithm of the scale factor in equation~\eqref{effective_action_in_a_g}. The cancellation was explicitly shown for scalar perturbations in \cite{senatore}, but although expected, it had not been shown for tensor perturbations until now. This is, in fact, a rather generic phenomenon for one-loop calculations involving light fields in de Sitter and a similar cancellation in a slightly different context had also been emphasized in \cite{perez-nadal08b}.

In contrast with existing computations of one-loop corrections to cosmological perturbations, where only equal times were considered, we have obtained the correlation function for arbitrary values of the two times. This is important so that one can check the full de Sitter invariance of the correlation function (the condition mentioned in the previous paragraph is a necessary but not sufficient condition for de Sitter invariance) since general de Sitter transformations do not preserve the equal-time condition. Furthermore, whereas calculations so far have only been approximate and typically valid for sufficiently late times, so that the co-moving mode under consideration is well outside the horizon, our calculation is exact and valid for arbitrary times, which makes it possible to check exactly whether de Sitter invariance is preserved by one-loop corrections in our case.
Moreover, since we considered arbitrary pairs of times and explicitly performed the regularization and renormalization of the UV divergences arising in our one-loop calculation by using dimensional regularization and introducing suitable local counter-terms in the bare gravitational action, we could explicitly see that although the renormalized two-point function is a well defined distribution, the coincident-time limit in spatial Fourier space diverges. (In position space, however, the equal-time limit is finite for spatially separate points.) That means that, strictly speaking, when loop corrections are included, the power spectrum is no longer a well-defined quantity in general.\footnote{This point had not been appreciated in previous calculations because all UV divergences in the power spectrum were removed by hand rather than through a detailed process of regularization and renormalization.}
A more detailed discussion on this subtle point is provided in appendix~\ref{appendix_coincidence}.

Some earlier calculations of loop corrections from conformal fields to cosmological perturbations (see \cite{wuetal11} and references therein) found terms in the result for the two-point function which are proportional to the initial time $t_0$ and diverge as one takes the limit $t_0 \to -\infty$. This would imply a bound on the number of e-foldings of inflation, established by requiring that the amplitude of primordial cosmological perturbations does not exceed the limits imposed by CMB and large-scale structure observations \cite{wuetal11}. Moreover, the need to consider a finite value of $t_0$ would prevent de Sitter invariance since the terms with explicit dependence on $t_0$ are incompatible, for fixed $t_0$, with the rescaling symmetry mentioned above (in our calculation instead any terms with dependence on $t_0$ disappear when properly taking the limit $t_0 \to -\infty$).
This question was discussed in detail in section~\ref{secular_terms}, where these problematic terms were identified as secular terms that arise due to the cancellation of the exponents of oscillatory factors when calculating the partial contribution to the two-point function often referred to as \emph{induced} fluctuations. Furthermore, we also explained on general grounds why such cancellation of oscillatory factors and the corresponding secular terms should not be present in the result of a perturbative CTP calculation of the full correlation function, a point confirmed at one-loop by our explicit result of section~\ref{twopoint_oneloop}. Finally, we clarified that considering only the contribution of the induced fluctuations without also taking into account the one-loop correction to the evolution of the \emph{intrinsic} fluctuations is the crucial reason for the lack of cancellation of the secular terms in \cite{wuetal11}.
It should be pointed out that other quantities (different from quantum correlation functions calculated using the approach described in section~\ref{overview}) can exhibit secular terms. However, that does not necessarily mean that those quantities can grow significantly: it may just signal a breakdown of perturbation theory for sufficiently long times, while a non-perturbative calculation would reveal a perfectly regular behavior. This situation is illustrated by the analysis in \cite{frv2011b}, where the evolution for arbitrary long times of linear perturbations of the semiclassical background around de Sitter (including effects of matter loops) is obtained by solving non-perturbatively the equation governing their dynamics.

We close this section briefly describing our plans for future work in order to extract further relevant information from the results obtained here and to generalize them.
One of our goals is to calculate the one-loop Riemann correlation function \cite{frv2012a}, which includes contributions from scalar, vectorial and tensorial metric perturbations. The tensorial contribution is the technically and conceptually most elaborate one, and the results obtained here are the key ingredient for achieving our goal. The Riemann correlation function (with appropriately raised indices) exhibits a number of interesting features. First, it is gauge-invariant even when the corrections due to matter loops (but not to graviton loops) are included. Second, it is compatible with de Sitter invariance (in contrast with other gauge-invariant quantities more commonly employed in cosmology). In fact, whether de Sitter invariance is preserved can be manifestly seen by expressing the result in terms of maximally symmetric bitensors \cite{allen1986,perez-nadal10}. So far the rescaling symmetry of our result mentioned above is a necessary but not sufficient condition for de Sitter invariance since it only corresponds (together with the $\mathrm{E}(3)$ symmetry of the spatially flat sections) to a subgroup of the full isometry group, $\mathrm{O}(4,1)$. Moreover, the additional isometries act nontrivially on the tensorial structure and the use of maximally symmetric bitensors becomes particularly convenient.
Thirdly, the Riemann tensor provides a complete and suitable characterization of the local geometry (geometrical observables associated with a finite region of spacetime can be obtained from it) while being insensitive to potentially IR divergent super-horizon contributions which cannot be probed by observables involving finite spacetime regions.

A particularly interesting generalization of our work is to consider massless minimally coupled fields instead of conformal ones since they typically give rise to much larger IR effects in de Sitter, and can even provide some hints on the viability of the proposed mechanism for a secular screening of the cosmological constant in the pure gravity case \cite{tsamis97,tsamis98}.
Having considered the conformal case first has, nevertheless, provided a good testing ground for our techniques. Furthermore, since it is a UV issue, the conclusion that the one-loop correction to the power spectrum is strictly speaking divergent, as discussed in appendix~\ref{appendix_coincidence}, applies in general (including the minimal coupling case). Similarly the cancellation of the secular terms discussed in section~\ref{twopoint_oneloop} also holds generically because its main contribution comes from modes well within the horizon. In this respect, an additional contribution of our work is to shed light on the claim of \cite{wuetal11} that even loops of conformal fields can give rise to large effects on cosmological perturbations.

\acknowledgments

M.~F.\ acknowledges financial support through an APIF scholarship from the Universitat de Barcelona and a FPU scholarship no.~AP2010-5453.
E.~V.\ and M.~F.\ also acknowledge partial financial support by the Research Projects MCI~FPA2007-66665-C02-02, FPA2010-20807-C02-02, CPAN~CSD2007-00042, with\-in the program Consolider-Ingenio~2010, and AGAUR~2009-SGR-00168.
During the final stages of this project, A.~R.\ was supported by the Deutsches Zentrum für Luft- und Raumfahrt (DLR) with funds provided by the Bundesministerium für Wirtschaft und Technologie under grant number DLR~50~WM~1136.

\appendix

\section{Metric expansion}
\label{appendix_metric}

For the perturbed metric $g_{ab}$, we have
\begin{equation}
\begin{split}
g_{ab} &= \eta_{ab} + \kappa h_{ab} \\
g^{ab} &= \eta^{ab} - \kappa h^{ab} + \kappa^2 h^a_m h^{bm} + \bigo{h^3} \\
h &= \eta^{ab} h_{ab} \\
\sqrt{-g} &= 1 + \frac{1}{2} \kappa h + \frac{1}{8} \kappa^2 h^2 - \frac{1}{4} \kappa^2 h_{mn} h^{mn} + \bigo{h^3} \eqend{.}
\end{split}
\end{equation}
where $\kappa^2 = 16 \mathpi G_\text{N}$ and all indices are raised and lowered with the unperturbed metric $\eta_{ab}$.

For the calculation of the Christoffel symbols (the connection) and the curvature tensors we will regard $h_{ab}$ as a tensor field in flat space. Moreover, we use $\dalembert = \eta^{mn} \partial_m \partial_n$ and $\partial^m = \eta^{mn} \partial_n$. Taking that into account, for the Christoffel symbols we get
\begin{equation}
\label{christoffel}
\begin{split}
\christoffel{a}{b}{c} &= \frac{1}{2} \kappa {S^a}_{bc} - \frac{1}{2} \kappa^2 h^a_m {S^m}_{bc} + \bigo{h^3} \\
{S^a}_{bc} &= \partial_b h^a_c + \partial_c h^a_b - \partial^a h_{bc} \\
{S^a}_{ac} &= \partial_c h \\
\partial_c h_{mb} &= \frac{1}{2} \eta_{an} \left( \delta^n_m {S^a}_{bc} + \delta^n_b {S^a}_{mc} \right) \\
\end{split}
\end{equation}
The calculation of the curvature tensors is then a simple exercise,
\begin{equation}
\begin{split}
{R^a}_{bcd} &= \kappa \partial_{[c} {S^a}_{d]b} - \kappa^2 h^a_m \partial_{[c} {S^m}_{d]b} - \frac{1}{2} \kappa^2 \eta_{pm} \eta^{aq} {S^p}_{q[c} {S^m}_{d]b} + \bigo{h^3} \\
R_{ab} &= \frac{1}{2} \kappa \left( \partial_m {S^m}_{ab} - \partial_a \partial_b h \right) - \kappa^2 h^n_m \partial_{[n} {S^m}_{b]a} - \frac{1}{2} \kappa^2 \eta_{mn} \eta^{cd} {S^n}_{d[c} {S^m}_{b]a} + \bigo{h^3} \\
R &= \kappa \left( \partial_m \partial_n h^{mn} - \dalembert h \right) + \kappa^2 h^{mn} \left( \partial_n \partial_m h + \dalembert h_{mn} - 2 \partial_n \partial^a h_{ma} \right) \\
&\quad- \frac{1}{4} \kappa^2 \left( 2 \partial_c h^{nc} - \partial^n h \right) \left( 2 \partial^a h_{na} - \partial_n h \right) + \frac{1}{4} \kappa^2 \left( 3 \partial_c h_{md} - 3 \partial_m h_{cd} \right) \left( \partial^c h^{md} \right) \\
&\quad+ \frac{1}{4} \kappa^2 \left( \partial_d h_{mc} \right) \left( \partial^c h^{md} \right) + \bigo{h^3} \eqend{.}
\end{split}
\end{equation}
Finally the Weyl tensor is given in four dimensions by
\begin{equation}
\label{weyl}
C_{abcd} = R_{abcd} - \frac{1}{2} \left( R_{ac} g_{bd} - R_{ad} g_{bc} + R_{bd} g_{ac} - R_{bc} g_{ad} \right) + \frac{1}{6} R ( g_{ac} g_{bd} - g_{ad} g_{bc} ) = \bigo{h} \eqend{.}
\end{equation}

\section{Conformal transformation}
\label{appendix_conformal}

Under the conformal transformation
\begin{equation}
\tilde{g}_{ab} = \mathe^{2\omega} g_{ab} \eqend{,}
\end{equation}
the transformed Christoffel symbols are given by
\begin{equation}
\varchristoffel{a}{b}{c} = \christoffel{a}{b}{c} + \left( \delta^a_b \delta^m_c + \delta^a_c \delta^m_b - g_{bc} g^{am} \right) \partial_m \omega \eqend{,}
\end{equation}
and the curvature tensors follow as
\begin{equation}
\begin{split}
{\tilde{R}^a}{}_{bcd} &= {R^a}_{bcd} + 4 g^{ak} \delta^m_{[c} g_{d][k} \left[ \nabla_{b]} \nabla_m \omega - ( \nabla_{b]} \omega ) ( \nabla_m \omega ) \right] - 2 \delta^a_{[c} g_{d]b} ( \nabla^m \omega ) ( \nabla_m \omega ) \\
\tilde{R}_{bd} &= R_{bd} - 2 \left[ \nabla_b \nabla_d \omega - ( \nabla_b \omega ) ( \nabla_d \omega ) + g_{bd} ( \nabla^m \omega ) ( \nabla_m \omega ) \right] - g_{bd} \nabla^m \nabla_m \omega \\
\mathe^{2\omega} \tilde{R} &= R - 6 \nabla^m \nabla_m \omega - 6 ( \nabla^m \omega ) ( \nabla_m \omega ) \eqend{,}
\end{split}
\end{equation}
where the covariant derivative $\nabla$ refers to the metric $g_{ab}$.

\section{Special functions}
\label{appendix_special}

We define the entire function $\Ein(z)$ by
\begin{equation}
\label{expintegral_ein_definition}
\Ein(z) = \int_0^z \frac{\mathe^t - 1}{t} \total t = \sum_{k=1}^\infty \frac{z^k}{k \, k!} \eqend{.}
\end{equation}
Its asymptotic expansion at infinity ($r \to \infty$) is given by
\begin{equation}
\Ein \left( \alpha r + \beta \right) \sim - \gamma -  \ln \left( - ( \alpha r + \beta ) \right) + \mathe^{\alpha r + \beta} \left[ \frac{1}{\alpha r}  + \frac{1-\beta}{\alpha^2 r^2} + \bigo{r^{-3}} \right] \eqend{,}
\end{equation}
where $\gamma$ is the Euler-Mascheroni constant.

The following useful integrals involving $\Ein(z)$ can be easily obtained by partial integration:
\begin{equation}
\begin{split}
\int \frac{\mathe^{a t}}{b t + c} \total t &= \frac{1}{b} \mathe^{-\tfrac{a c}{b}} \left[ \Ein\left[ \frac{a}{b} \left( b t + c \right) \right] + \ln \left( b t + c \right) \right] \\
\int \mathe^{a t} \ln ( b t + c ) \total t &= \frac{1}{a} \left[ \left( \mathe^{a t} - \mathe^{-\tfrac{a c}{b}} \right) \ln \left( b t + c \right) - \mathe^{-\tfrac{a c}{b}} \Ein\left[ \frac{a}{b} \left( b t + c \right) \right] \right] \\
\int \mathe^{a t} \Ein ( b t + c ) \total t &= \frac{1}{a} \mathe^{a t} \Ein ( b t + c ) + \frac{1}{a} \mathe^{-\tfrac{a c}{b}} \left[ \Ein\left( \frac{a}{b} \left( b t + c \right) \right) - \Ein\left( \frac{a+b}{b} \left( b t + c \right) \right) \right] \eqend{.}
\end{split}
\end{equation}
These integrals depend continuously on the parameters $a$ and $c$, so that for instance the integral of $\Ein (b t + c)$ can be calculated by taking the limit $a \to 0$ on the right side of the last integral.

\section{Distributions and their Fourier transforms}
\label{appendix_distributions}

Here we analyze the main distributions which appear in section~\ref{twopoint_oneloop}. Good references for the general theory of distributions are \cite{schwartz} and \cite{gelfand}.

The principal value distribution, denoted by $\mathcal{P} \dfrac{1}{t}$, is defined as
\begin{equation}
\mathcal{P} \frac{1}{t} = \distlim_{\epsilon \to 0} \frac{\Theta(t-\epsilon) + \Theta(-t-\epsilon)}{t} \eqend{,}
\end{equation}
where $\distlim$ stands for ``limit in the sense of distributions'', i.e.
\begin{equation}
\int \mathcal{P} \frac{1}{t} f(t) \total t = \lim_{\epsilon \to 0} \left[ \int \frac{\Theta(t-\epsilon) + \Theta(-t-\epsilon)}{t} f(t) \total t \right] \eqend{,}
\end{equation}
where $f(t)$ is an appropriate (fast decaying) test function. The Fourier transform can be calculated to give
\begin{equation}
\label{principal_value}
\begin{split}
\int \mathcal{P} \frac{1}{t} \mathe^{\mathi \omega t} \total t &= \lim_{\epsilon \to 0} \left[ \left( \int_{-\infty}^{-\epsilon} + \int_\epsilon^\infty \right) \frac{\mathe^{\mathi \omega t}}{t} \total t \right] = \lim_{\epsilon \to 0} \left[ \lim_{r \to \infty} \int_{-r}^{-\epsilon} \frac{\mathe^{\mathi \omega t}}{t} \total t + \lim_{r \to \infty} \int_\epsilon^r \frac{\mathe^{\mathi \omega t}}{t} \total t \right] \\
&= \lim_{\epsilon \to 0} \left[ \lim_{r \to \infty} \left[ \Ein\left( \mathi \omega t \right) + \ln \abs{t} \right]_{-r}^{-\epsilon} + \lim_{r \to \infty} \left[ \Ein\left( \mathi \omega t \right) + \ln \abs{t} \right]_\epsilon^r \right] \\
&= \lim_{\epsilon \to 0} \left[ \Ein\left( - \mathi \omega \epsilon \right) + \ln ( \mathi \omega ) - \ln ( - \mathi \omega ) - \Ein\left( \mathi \omega \epsilon \right) \right] \\
&= \mathi \arg ( \mathi \omega ) - \mathi \arg ( - \mathi \omega ) = \mathi \mathpi \sgn \omega \eqend{.}
\end{split}
\end{equation}

Another distribution of interest is $\mathcal{P}' \dfrac{\Theta(t)}{t}$, which is defined as
\begin{equation}
\label{def_distri_pf_thetat}
\mathcal{P}' \frac{\Theta(t)}{t} = \distlim_{\epsilon \to 0} \left[ \frac{\Theta(t-\epsilon)}{t} + \delta(t) ( \ln ( \mu_0 \epsilon ) + \gamma ) \right] \eqend{,}
\end{equation}
where $\mu_0 > 0$ is an arbitrary reference energy scale to make the definition dimensionally correct, and we use the symbol $\mathcal{P}'$ to distinguish from the previous $\mathcal{P}$ which did not include a $\delta$ distribution. Its Fourier transform can be calculated by proceeding as in the previous case, and is given by
\begin{equation}
\begin{split}
\int \mathcal{P}' \frac{\Theta(t)}{t} \mathe^{\mathi \omega t} \total t &= \lim_{\epsilon \to 0} \left[ \int_\epsilon^\infty \frac{\mathe^{\mathi \omega t}}{t} \total t + \ln ( \mu_0 \epsilon ) + \gamma \right] \\
&= \lim_{\epsilon \to 0} \lim_{r \to \infty} \left[ [ - \gamma - \ln ( - \mathi \omega ) - \ln r ] + \ln r - \Ein\left( \mathi \omega \epsilon \right) - \ln \epsilon + \ln \epsilon + \gamma + \ln \mu_0 \right] \\
&= \ln \left( \frac{\mu_0}{\abs{\omega}} \right) + \frac{\mathi \mathpi}{2} \sgn \omega \eqend{.}
\end{split}
\end{equation}
Using those two results we may write the following compact expression:
\begin{equation}
\label{log_omega_fourier}
\begin{split}
- \int \ln \left( \frac{\omega}{\mu_0} \right)^2 \mathe^{-\mathi \omega t} \frac{\total \omega}{2 \mathpi} &= 2 \mathcal{P}' \frac{\Theta(t)}{t} - \mathcal{P} \frac{1}{t} \\
&= \distlim_{\epsilon \to 0} \left[ \frac{\Theta(\abs{t}-\epsilon)}{\abs{t}} + 2 \delta(t) ( \ln ( \mu_0 \epsilon ) + \gamma ) \right] \eqend{,}
\end{split}
\end{equation}
where we denote the distribution defined by the last line as $\mathcal{P}' \abs{t}^{-1}$. Simply applying to this equation the formula for the (inverse) Fourier transform of a function with the argument shifted by a constant, one gets
\begin{equation}
\label{log_omega_fourier2}
- \int \ln \left( \frac{a+\omega}{\mu_0} \right)^2 \mathe^{- \mathi \omega t} \frac{\total \omega}{2 \mathpi} = \mathe^{\mathi a t} \mathcal{P}' \frac{1}{\abs{t}} \eqend{.}
\end{equation}

We have now all the elements for calculating the temporal inverse Fourier transform of
\begin{equation}
\ln \left( \frac{p^2 \pm \mathi 0}{\bar{\mu}^2} \right) = \ln \abs{\frac{p^2}{\bar{\mu}^2}} \pm \mathi \mathpi \Theta(-p^2) \eqend{.}
\end{equation}
First, from equation~\eqref{log_omega_fourier2} we obtain
\begin{equation}
\label{ln_distri_ft1}
\begin{split}
\int \ln \abs{\frac{p^2}{\bar{\mu}^2}} \mathe^{- \mathi p^0 t} \frac{\total p^0}{2 \mathpi}
&= \int \left[ \ln \abs{p^0 + \abs{\vec{p}}} + \ln \abs{p^0 - \abs{\vec{p}}} - 2 \ln \bar{\mu}
\right] \\
&= - \cos \left( \abs{\vec{p}} t \right) \mathcal{P}' \frac{1}{\abs{t}} - \delta(t) \ln \left( \frac{\bar{\mu}^2}{\mu_0^2} \right) \eqend{.}
\end{split}
\end{equation}
Second, writing $\Theta(-p^2)$ as $1-\Theta \big( \abs{\vec{p}}^2 - (p^0)^2 \big)$, we can immediately calculate
\begin{equation}
\label{ln_distri_ft2}
\int \Theta(-p^2) \mathe^{- \mathi p^0 t} \frac{\total p^0}{2 \mathpi} = \left[ \delta(t) - \frac{\sin \left( \abs{\vec{p}} t \right)}{\mathpi t} \right] \eqend{.}
\end{equation}
Combining the two we finally get
\begin{equation}
\label{ln_distri_ft}
\int \ln \left( \frac{p^2 \pm \mathi 0}{\bar{\mu}^2} \right) \mathe^{- \mathi p^0 t} \frac{\total p^0}{2 \mathpi} = - \cos \left( \abs{\vec{p}} t \right) \mathcal{P}' \frac{1}{\abs{t}} - \delta(t) \ln \left( \frac{\bar{\mu}^2}{\mu_0^2} \right) \pm \mathi \mathpi \left[ \delta(t) - \frac{\sin \left( \abs{\vec{p}} t \right)}{\mathpi t} \right] \eqend{,}
\end{equation}
which provides the kernels \eqref{kernels_fts} which we need for our calculation.

\section{Calculation of the integrals in equation \texorpdfstring{\protect\eqref{propagator_at_first_order_two}}{(4.26)}}
\label{appendix_integrals}

Here we will explain in some detail how to calculate the integrals in equation~\eqref{propagator_at_first_order} using the $\mathi\epsilon$ prescription described in section~\ref{overview_inin}.

The first thing that we will do is to substitute the definitions of the distributions~\eqref{kernels_fts} from appendix~\ref{appendix_distributions} into these integrals. We then note that the $\delta(t) \ln(\mu_0)$ part from equation~\eqref{def_distri_pf_thetat} appearing in the kernel $\tilde{L}$ can be used to integrate over $\tau'$, so that the logarithm can be combined with the one appearing in the third integral of equation~\eqref{propagator_at_first_order}. Setting $\mu_0 = \bar{\mu}$, this gives
\begin{equation}
\label{explicit_integral}
\int \left[ ( \partial^2_\tau + \vec{p}^2 ) \tilde{G}^0(\eta, \tau, \vec{p}) \right] \begin{pmatrix} 1 & 0 \\ 0 & -1 \end{pmatrix} \left[ ( \partial^2_\tau + \vec{p}^2 ) \tilde{G}^0(\tau, \eta', \vec{p}) \right] \ln \left( \frac{\bar{\mu}}{H} \right) \total \tau \eqend{.}
\end{equation}
We will perform this integral explicitly, which will serve as an illustration of the computation of the other integrals since they all follow the same general pattern.

The two-point function that we will calculate is the imaginary unit times the positive Wightman function $\tilde{G}^{-+}$ defined in~\eqref{propagator_functional_derivative}. The contribution of the above integral is then given by
\begin{equation}
- 3 \mathi \alpha \kappa^2 \ln \left( \frac{\bar{\mu}}{H} \right) J \eqend{,}
\end{equation}
where
\begin{equation}
\label{j_integral}
\begin{split}
J &= \int_{t_0^+}^T \left[ ( \partial^2_\tau + \vec{p}^2 ) \tilde{G}^0_{-+}(\eta, \tau, \vec{p}) \right] \left[ ( \partial^2_\tau + \vec{p}^2 ) \tilde{G}^0_{++}(\tau, \eta', \vec{p}) \right] \total \tau \\
&\quad- \int_{t_0^-}^T \left[ ( \partial^2_\tau + \vec{p}^2 ) \tilde{G}^0_{--}(\eta, \tau, \vec{p}) \right] \left[ ( \partial^2_\tau + \vec{p}^2 ) \tilde{G}^0_{-+}(\tau, \eta', \vec{p}) \right] \total \tau \eqend{.}
\end{split}
\end{equation}
We are integrating from the initial time $t_0^\pm$ to the final time $T$, according to whether the integration variable $\tau$ belongs to the upper or lower half of the contour shown in figure \ref{ctp_path}.

For the Wightman functions we have
\begin{equation}
\label{wightman_functions_derivatives}
\begin{split}
( \partial^2_\tau + \vec{p}^2 ) \tilde{G}^0_{-+/+-}(\eta, \tau, \vec{p}) &= \frac{2}{\tau} \partial_\tau \tilde{G}^0_{-+/+-}(\eta, \tau, \vec{p}) \\
( \partial^2_\tau + \vec{p}^2 )^2 \tilde{G}^0_{-+/+-}(\eta, \tau, \vec{p}) &= 0 \eqend{,}
\end{split}
\end{equation}
since these functions are proportional to $f$ or $f^*$, which are solutions of the equation of motion~\eqref{equation_order_zero_f}. To make sure that we do not get a contribution from the arbitrary final time $T$, we first write
\begin{equation}
\begin{split}
\tilde{G}^0_{++}(\eta, \eta', \vec{p}) &= \Theta(\eta-\eta') \left[ \tilde{G}^0_{-+}(\eta, \eta', \vec{p}) - \tilde{G}^0_{+-}(\eta, \eta', \vec{p}) \right] + \tilde{G}^0_{+-}(\eta, \eta', \vec{p}) \\
&= \tilde{G}^0_\text{ret}(\eta, \eta', \vec{p}) + \tilde{G}^0_{+-}(\eta, \eta', \vec{p}) \eqend{,} \\
\tilde{G}^0_{--}(\eta, \eta', \vec{p}) &= - \tilde{G}^0_\text{ret}(\eta, \eta', \vec{p}) + \tilde{G}^0_{-+}(\eta, \eta', \vec{p}) \eqend{,}
\end{split}
\end{equation}
which simplifies $J$ to
\begin{equation}
\begin{split}
J &= \int_{t_0^+}^T \left[ ( \partial^2_\tau + \vec{p}^2 ) \tilde{G}^0_{-+}(\eta, \tau, \vec{p}) \right] \left[ ( \partial^2_\tau + \vec{p}^2 ) \tilde{G}^0_\text{ret}(\eta', \tau, \vec{p}) \right] \total \tau \\
&\quad+ \int_{t_0^-}^T \left[ ( \partial^2_\tau + \vec{p}^2 ) \tilde{G}^0_\text{ret}(\eta, \tau, \vec{p}) \right] \left[ ( \partial^2_\tau + \vec{p}^2 ) \tilde{G}^0_{-+}(\tau, \eta', \vec{p}) \right] \total \tau \\
&\quad- \int_{t_0^-}^{t_0^+} \left[ ( \partial^2_\tau + \vec{p}^2 ) \tilde{G}^0_{-+}(\eta, \tau, \vec{p}) \right] \left[ ( \partial^2_\tau + \vec{p}^2 ) \tilde{G}^0_{-+}(\tau, \eta', \vec{p}) \right] \total \tau \eqend{.}
\end{split}
\end{equation}
We see that the result will not depend on $T$ provided that $T>\eta,\eta'$, as expected. Next, we calculate
\begin{equation}
\begin{split}
\partial_\tau \tilde{G}^0_\text{ret}(\eta, \tau, \vec{p}) &= \Theta(\eta - \tau) \left( \partial_\tau \tilde{G}^0_{-+}(\eta, \tau, \vec{p}) - \partial_\tau \tilde{G}^0_{+-}(\eta, \tau, \vec{p}) \right) \\
( \partial^2_\tau + \vec{p}^2 ) \tilde{G}^0_\text{ret}(\eta, \tau, \vec{p}) &= \frac{2}{\tau} \partial_\tau \tilde{G}^0_\text{ret}(\eta, \tau, \vec{p}) - H^2 \eta^2 \delta(\tau - \eta) \eqend{,}
\end{split}
\end{equation}
and together with equation~\eqref{wightman_functions_derivatives} this leads to
\begin{equation}
\begin{split}
J &= - 2 H^2 \eta \left[ \partial_\eta \tilde{G}^0_{-+}(\eta, \eta', \vec{p}) \right] - 2 H^2 \eta' \left[ \partial_{\eta'} \tilde{G}^0_{-+}(\eta, \eta', \vec{p}) \right] \\
&\quad+ 4 \int_{t_0^+}^T \frac{1}{\tau^2} \left[ \partial_\tau \tilde{G}^0_{-+}(\eta, \tau, \vec{p}) \right] \left[ \partial_\tau \tilde{G}^0_\text{ret}(\eta', \tau, \vec{p})\right] \total \tau \\
&\quad+ 4 \int_{t_0^-}^T \frac{1}{\tau^2} \left[ \partial_\tau \tilde{G}^0_\text{ret}(\eta, \tau, \vec{p}) \right] \left[ \partial_\tau \tilde{G}^0_{-+}(\tau, \eta', \vec{p}) \right] \total \tau \\
&\quad- 4 \int_{t_0^-}^{t_0^+} \frac{1}{\tau^2} \left[ \partial_\tau \tilde{G}^0_{-+}(\eta, \tau, \vec{p}) \right] \left[ \partial_\tau \tilde{G}^0_{-+}(\tau, \eta', \vec{p}) \right] \total \tau \eqend{.}
\end{split}
\end{equation}
These integrals can now be easily computed. Using the definition of the Wightman functions from equation~\eqref{propagators_at_order_zero}, we obtain
\begin{equation}
\begin{split}
J &= - \frac{H^4}{\abs{\vec{p}}} (\eta')^2 ( \abs{\vec{p}} \eta - \mathi ) \mathe^{- \mathi \abs{\vec{p}} (\eta-\eta')} + \frac{H^4}{\abs{\vec{p}}} \eta^2 ( \abs{\vec{p}} \eta' + \mathi ) \mathe^{- \mathi \abs{\vec{p}} (\eta-\eta')} \\
&\quad- \mathi \frac{H^4}{2 \abs{\vec{p}}^3} ( \abs{\vec{p}} \eta - \mathi ) ( \abs{\vec{p}} \eta' - \mathi ) \mathe^{- \mathi \abs{\vec{p}} (\eta+\eta')} \left( \mathe^{2 \mathi \abs{\vec{p}} \eta'} - \mathe^{2 \mathi \abs{\vec{p}} t_0^+} \right) \\
&\quad- \mathi \frac{H^4}{2 \abs{\vec{p}}^3} ( \abs{\vec{p}} \eta + \mathi ) ( \abs{\vec{p}} \eta' + \mathi ) \mathe^{\mathi \abs{\vec{p}} (\eta+\eta')} \left( \mathe^{- 2 \mathi \abs{\vec{p}} \eta} - \mathe^{- 2 \mathi \abs{\vec{p}} t_0^-} \right) \\
&\quad+ \frac{H^4}{\abs{\vec{p}}^2} ( \abs{\vec{p}} \eta - \mathi ) ( \abs{\vec{p}} \eta' + \mathi ) \mathe^{- \mathi \abs{\vec{p}} (\eta-\eta')} \left( \eta' - \eta \right) \eqend{.}
\end{split}
\end{equation}
Finally, we take the limit $t_0^\pm \to -\infty(1 \mp \mathi\epsilon)$, which causes the exponentials containing $t_0^\pm$ to vanish. After collecting terms we get
\begin{equation}
J = \mathi \frac{H^4}{\abs{\vec{p}}^3} \left( \abs{\vec{p}} \eta - \mathi \right) \left( \abs{\vec{p}} \eta' + \mathi \right) \mathe^{- \mathi \abs{\vec{p}} (\eta-\eta')} = 2 \mathi f(\eta, \eta', \vec{p}) \eqend{,}
\end{equation}
so that the contribution of the integral \eqref{explicit_integral} is
\begin{equation}
6 \alpha \kappa^2 \ln \left( \frac{\bar{\mu}}{H} \right) f(\eta, \eta', \vec{p})
\end{equation}
as given in the final result~\eqref{two_point_function_h}.

For the remaining integrals containing the kernels \eqref{def_kernels_zwei}, we need the special function $\Ein$, which is given (together with the necessary indefinite integrals) in appendix \ref{appendix_special}. To further facilitate the calculation, it is convenient to combine the integrals containing the kernels $\tilde{L}$ and $\tilde{N}$ as well as $\tilde{D}$ and $\tilde{N}$ to give the distributions
\begin{equation}
\mathe^{\pm \mathi \abs{\vec{p}} (\tau-\tau')} \mathcal{P} \frac{1}{\tau-\tau'} \eqend{,}
\end{equation}
which allows to take the limit $t_0^\pm \to -\infty(1 \mp \mathi\epsilon)$ already in the result of the integral over $\tau'$ before integrating over $\tau$.

\section{Coincidence-limit divergence}
\label{appendix_coincidence}

The one-loop two-point function of the metric perturbations obtained in section~\ref{twopoint_oneloop} is a well-defined spacetime distribution, which can be obtained by computing the inverse Fourier transform of equation~\eqref{propagator_at_first_order}. This means that when integrated with appropriate spacetime test functions, it will give a finite result. However, if one takes the equal-time limit $\eta' \to \eta$, the resulting two-point function is no longer well defined as a distribution\footnote{Mathematically, this corresponds to the fact that given a distribution in a manifold, its restriction to a submanifold is in general not guaranteed to be a well-defined distribution. The restriction should be implemented by considering a sequence of test functions on the full manifold with decreasing width around the submanifold, applying the distribution to the test functions and seeing if the limit of the resulting sequence (the zero-width limit) is well-defined.} and its spatial Fourier transform, which corresponds to the power spectrum, is divergent. It should be emphasized that the usual divergences arising in our one-loop calculation have already been properly regulated and canceled out by local counterterms in the bare gravitational action. There is nothing else to be subtracted: the two-point function is a well-defined distribution and it will give finite results when employed to calculate proper physical observables. It is just that when including loop corrections, the power spectrum is strictly speaking not a well-defined observable any more.

We elaborate below on this point, which might sound surprising at first, and illustrate it with a simple example in flat space that shares all the key features. The fact that this issue also appears in flat space is not unexpected since it is a UV effect.

\subsection{Simple example in flat space}

Let us consider the following connected two-point function for a massless real field $\hat{\phi}(x)$ in the Minkowski vacuum:
\begin{equation}
\label{noise}
\frac{1}{8 \mathpi^2} N(x,x') = \frac{1}{2} \left\langle \big\{ \hat{\phi}^2(x),  \hat{\phi}^2(x') \big\} \right\rangle_\text{c} \eqend{,}
\end{equation}
where we introduced the notation $\expect{ \hat{A}(x) \hat{B}(x') }_\text{c} = \expect{\hat{A}(x) \hat{B}(x')} - \expect{\hat{A}(x)} \expect{\hat{B}(x')}$. It corresponds to the real part of the product of two Wightman functions $-2 \Re \left[ G^+(x,x') \right]^2$. One can calculate it in (full) Fourier space as a convolution over four-momentum \cite{martin00}, and the result is
\begin{equation}
\label{noise_fourier}
\bar{N}(p) = \mathpi \Theta(-p^2) \eqend{.}
\end{equation}
Note that we included a factor $(8\mathpi^2)^{-1}$ in equation~\eqref{noise}, so that the kernel $N(x)$ coincides with the one introduced in equations~\eqref{def_kernels_zwei}.
The expression in spacetime coordinates is then obtained by calculating the inverse Fourier transform and is given by
\begin{equation}
\label{hadamard}
N(x-x') = \frac{1}{\mathpi^2} \, \mathcal{P} \frac{1}{\big( (x-x')^2 \big)^2} \eqend{,}
\end{equation}
where $\mathcal{P}$ denotes Hadamard's finite part distribution \cite{schwartz}. Hence, we can explicitly see that $N(x-x')$ is a well-defined spacetime distribution and its Fourier transform is finite.
Alternatively one can consider the spatial Fourier transform
\begin{equation}
\label{noise_k}
\tilde{N}(\eta - \eta',\vec{p}) = - \frac{\sin \left[ \abs{\vec{p}} (\eta-\eta') \right]}{\eta - \eta'} + \mathpi \delta(\eta-\eta') \eqend{,}
\end{equation}
which is again a well-defined distribution, but diverges in the equal time limit $\eta' \to \eta$, since it has singular support there. Hence, the power spectrum associated to $N(x,x')$ is divergent. In contrast, the Fourier transform with respect to time is given by
\begin{equation}
\label{noise_omega}
\breve{N} (\omega, \vec{x}-\vec{x}') = 8 \mathpi^2 \frac{\omega}{r^2} \left( \frac{\sin(\omega r)}{\omega r} -\cos(\omega r) \right) \eqend{,}
\end{equation}
where $r = \abs{\vec{x}-\vec{x}'}$, and its spatial coincidence limit $N(\omega, \mathbf{0}) = \omega^3/3$ is regular.

There is, in fact, a relatively simple qualitative explanation for the divergence of the power spectrum associated with the product of two Wightman functions as well as for the radically different behavior of the temporal and spatial Fourier transforms \cite{hu07}. The divergence of the equal time limit for the spatial Fourier transform (and thus the power spectrum) can be easily understood if one writes the product of Wightman functions as a convolution in 3-momentum space:
\begin{equation}
\label{convolution1}
\int \tilde{G}^+(\eta-\eta', \vec{p}-\vec{q}) \tilde{G}^+(\eta-\eta', \vec{q}) \frac{\total^3 q}{(2\mathpi)^3} = - \int \frac{\mathe^{\mathi(\omega_{\vec{p}-\vec{q}} + \omega_{\vec{q}})(\eta-\eta')}} {4\, \omega_{\vec{p}-\vec{q}} \, \omega_{\vec{q}}} \frac{\total^3 q}{(2\mathpi)^3} \eqend{,}
\end{equation}
where $\omega_\vec{p} = \abs{\vec{p}}$. For $\eta = \eta'$ power counting immediately reveals that the integral diverges. Note that momenta with arbitrarily large modulus $\abs{\vec{q}}$ contribute to the momentum integral even when keeping the total momentum $\vec{p}$ fixed, because they appear with opposite signs in the two Wightman functions. This is crucial to understanding the different behavior of the spatial and temporal Fourier transforms: although the Wightman function only involves positive frequencies, it admits both positive and negative spatial momenta along any direction. More specifically, one can write the product of the two Wightman functions as a convolution in full Fourier space\footnote{Equations~\eqref{convolution1} and \eqref{convolution2} can be easily generalized to the massive case by taking $\omega_\vec{p} = \sqrt{\vec{p}^2 + m^2}$ and adding $m^2$ to the arguments of the delta functions in equation~\eqref{convolution2}, so that for instance $q^2$ becomes $q^2 + m^2$.}
\begin{equation}
\label{convolution2}
\begin{split}
\left[ G^+(x,x') \right]^2 &= - \frac{1}{(2\mathpi)^6} \int \mathe^{\mathi p^0 (\eta-\eta')} \mathe^{-\mathi \vec{p} (\vec{x}-\vec{x}')} \int \delta(q^2) \delta((p-q)^2) \Theta(q^0) \Theta(p^0-q^0) \total^4 q \total p^0 \total^3 p \\
&= - \frac{1}{(2\mathpi)^6} \int \mathe^{\mathi p^0 (\eta-\eta')} \mathe^{-\mathi \vec{p} (\vec{x}-\vec{x}')} \int \frac{1}{2 \omega_{\vec{q}}} \delta((p-q)^2) \Theta(p^0 - \omega_\vec{q}) \total^3 q \total p^0 \total^3 p \eqend{,}
\end{split}
\end{equation}
where $q^0 = \omega_\vec{q} = \abs{\vec{q}}$ in the last equality. The spatial coincidence limit of the temporal Fourier transform corresponds to fixing $p^0$ and integrating over $\vec{p}$ the last integral in equation~\eqref{convolution2}. This double integral over $\vec{p}$ and $\vec{q}$ is finite because $\abs{\vec{q}}$ is bounded by the Heaviside function, which corresponds to the positive frequency condition of the Wightman function and implies $\omega_\vec{q} \leq p^0$.
In contrast, the equal time limit of the spatial Fourier transform amounts to fixing $\vec{p}$ and integrating over $p^0$ the last integral in equation~\eqref{convolution2}, which essentially eliminates the bound on $\abs{\vec{q}}$ imposed by the Heaviside function and leads to a divergent result due to contributions with arbitrarily large $\abs{\vec{q}}$. In fact, the result of integrating over $p^0$ corresponds exactly to the right-hand side of equation~\eqref{convolution1}.

We have considered the example of the correlation function \eqref{noise} in flat space for simplicity, but the essential points also hold for other correlation functions calculated at one-loop order in flat space. 
In fact, the anticommutator of the stress tensor can be obtained by applying differential operators to \eqref{noise} \cite{martin00}. For a conformal field its projection to the TT sector is given by
\begin{equation}
\label{noise_TT}
\frac{1}{2} \expect{ \big\{ \hat{T}_{ab} (x), \hat{T}_{cd} (x') \big\} }_\text{c}^\text{TT}
= \frac{1}{1920 \mathpi^2} \int \mathe^{\mathi \vec{p} (\vec{x} - \vec{x}')} P_{abcd} \, \left( \partial_\eta^2 + \abs{\vec{p}}^2 \right)^2 \tilde{N}(\eta-\eta', \vec{p}) \frac{\total^3 p}{(2\pi)^3} \eqend{,}
\end{equation}
where the TT projector $P_{abcd}$ has been defined in equation~\eqref{TTprojector}. Equation~\eqref{noise_TT} can be immediately extended to a $n$-dimensional spatially flat FLRW spacetime with scale factor $a(\eta)$ if one multiplies the flat space result by $a^{2-n}(\eta)\, a^{2-n}(\eta')$ \cite{martin99,eftekharzadeh11}.
Moreover, since it is a UV effect, one expects that the main conclusions in this appendix will also apply to Hadamard states in general curved spacetimes.

It should be stressed that equation~\eqref{noise_TT} enables one to anticipate easily
the singular coincidence limit of the power spectrum for the tensor metric perturbations even before calculating the time integrals corresponding to the interaction vertices in the Feynman diagram for the one-loop correction of the metric two-point function, as carried out in section~\ref{twopoint_oneloop}.
The argument is the following. First, one takes into account that the symmetrized connected two-point function of the Einstein tensor is given at one loop by the symmetrized connected two-point function of the stress tensor in the de Sitter background (the amputated loop diagram) \cite{perez-nadal10}. Note that such a two-point function is given by a well-defined spacetime distribution and it is not affected by the contributions from the renormalization counterterms. As mentioned in the previous paragraph, for conformal fields the stress tensor two-point function in a spatially flat FLRW metric is directly related to the result in Minkowski spacetime, whose projection to the tensorial part is given by equation~\eqref{noise_TT}. One can see that it contains terms with up to four time derivatives acting on the Dirac $\delta$ distribution function contributing to the kernel $\tilde{N}(\eta-\eta', \vec{p})$, as given by equation~\eqref{noise_k}. Since the two-point function of the Einstein tensor results from applying a fourth-order differential operator to the two-point function of the metric perturbations, one can conclude that the spatial Fourier transform of the latter should necessarily contain terms which diverge like a $\delta$ distribution in the equal-time limit.

One might try to circumvent the singular coincidence limit by defining the power spectrum not as in equation~\eqref{powerspectrum_def}, but as a constant times the quantum-corrected graviton mode functions derived from the semiclassical Einstein equations, along the lines of the treatment of Park and Woodard \cite{parkwoodard1,parkwoodard2} for the massless, minimally coupled scalar.

\subsection{Positivity of the power spectra}

Given a Hermitian operator $\hat{Q}$, its two-point function must be positive semidefinite since $\expect{\hat{Q}\hat{Q}} = \expect{\hat{Q}^\dagger \hat{Q}} \geq 0$. The same is true for the connected two-point function $\expect{\hat{Q}\hat{Q}}_\text{c}$ because for any given state $\hat{Q} - \expect{\hat{Q}}$ is also a Hermitian operator. In particular, one can consider the operator $\hat{\phi}^2(x)$ smeared with a suitable (rapidly decreasing) test function $F(x)$ as an example of such Hermitian operators:
\begin{equation}
\hat{Q} = \int F(x)\, \hat{\phi}^2(x) \total^4 x \eqend{.}
\label{smeared_op}
\end{equation}
(Note that although $\hat{\phi}^2(x)$ is a divergent composite operator, $\hat{\phi}^2(x) - \expect{\hat{\phi}^2(x)}$ is finite.)
From the argument above one can immediately conclude that the kernel $N(x,x')$ in equation~\eqref{noise} is positive semidefinite in the following sense:
\begin{equation}
S[F] = \iint F(x) N(x,y) F(y) \total^4 x \total^4 y = \iint \abs{\tilde{f}_s(\vec{p})}^2 f_{\eta_0}(\tau) \tilde{N}(\tau,\tau',\vec{p}) f_{\eta_0}(\tau') \total \tau \total \tau' \geq 0 \eqend{,}
\label{smeared_noise1}
\end{equation}
where we have assumed spatial translation invariance and that the test function factorizes into a temporal and a spatial part: $F(x) = f_s(\vec{x}) f_{\eta_0} (\eta)$, with $f_{\eta_0} (\eta)$ being some test function centered around $\eta_0$, such as
\begin{equation}
f_{\eta_0} (\eta) = \frac{1}{\sqrt{2\mathpi} \, \sigma_t} \mathe^{-\tfrac{(\eta-\eta_0)^2}{2\sigma_t^2}} \eqend{.}
\end{equation}
If we took the limit $\sigma_t \to 0$, we would be naively led to conclude that the associated power spectrum $\delta^2_{\vec{p}}(\eta) \sim \tilde{N}\left(\eta,\eta;\vec{p}\right)$ is positive semidefinite.

However, if one takes the simple flat space example given by equation~\eqref{noise} and neglects the local term, which diverges in the equal time limit, one gets $\delta^2_{\vec{p}}(\eta) \sim \tilde{N}\left(\eta,\eta,\vec{p}\right) = -\abs{\vec{p}} < 0$. This highlights the importance of the local term, which cannot be neglected in general. Indeed, if we keep the local term, we have
\begin{equation}
\label{smeared_noise2}
\frac{S[F]}{\abs{\tilde{f}_s(\vec{p})}^2} = - \iint f_{\eta_0}(\tau) \frac{\sin\left[\abs{\vec{p}}(\tau-\tau')\right]}{\tau-\tau'} f_{\eta_0}(\tau') \total \tau \total \tau' + \mathpi \int f_{\eta_0}^2(\tau) \total \tau \approx - \abs{\vec{p}} + \frac{\sqrt{\pi}}{2 \sigma_t} > 0 \eqend{,}
\end{equation}
where we considered the limit $\abs{\vec{p}} \sigma_t \ll 1$ in the last equality.

Similar conclusions apply in general to other two-point functions at one loop. An immediate example is the connected two-point function of the stress tensor for the Minkowski vacuum, whose transverse traceless projection is given by equation~\eqref{noise_TT} and to which the previous arguments can be straightforwardly extended by smearing with tensorial test fields $F^{ab}(x)$ instead of test functions $F(x)$. Furthermore, as already mentioned above, for a conformal field the connected stress tensor correlation function in $n$-dimensional flat space is simply related by a factor $a^{2-n}(\eta)\, a^{2-n}(\eta')$ to the result in an arbitrary spatially flat FLRW spacetime with scale factor $a(\eta)$. The correlation function of the metric perturbations, which we have calculated here perturbatively up to order $\kappa^2$ for tensor perturbations around de Sitter would be another example. In this case, the positivity requirement only implies a positive result for the tree-level contribution (of order $\kappa^0$) in general, since the perturbative expansion is only asymptotic, i.e.\ valid for $\kappa \to 0$. Nevertheless, if one considers correlation functions for other quantities that can be obtained from the metric two-point function but for which the tree level contribution vanishes, positivity enforces a nontrivial condition for the loop correction (of order $\kappa^2$) as the first term in this asymptotic expansion. One such example is the connected two-point function of the Ricci tensor (with one index raised), which vanishes at tree level for perturbations around a de Sitter background, and essentially coincides with the connected stress tensor correlation function.

\subsection{Physical implications}

In section~\ref{twopoint_oneloop} we found a divergent result for the renormalized one-loop correction to the power spectrum of tensor metric perturbations. As illustrated with simpler examples and explained with general arguments in the previous two subsections, this is in fact a fairly generic situation for one-loop corrections. Here we elaborate on the physical interpretation and implications of this fact.

As emphasized above, the spatial Fourier transform of the two-point function is a well-defined distribution in terms of the two times: it is only the coincidence limit which is divergent. Truly physical observables, nevertheless, will involve integrals over the two time arguments and will hence be finite.
Furthermore, although the underlying reason for the divergence of the power spectrum is of UV nature, involving points with small invariant separations, it can easily ``contaminate'' attempts to extract IR features if one is not careful enough. In particular, simply considering spatial Fourier modes with sufficiently long wavelength does not necessarily exclude UV effects (unless one smears over sufficiently long times as well): calculating the spatial Fourier transform for equal times involves integrating over points with arbitrarily small spacetime separations. Alternatively, this can be seen in Fourier space as a consequence of modes with spatial momenta of arbitrarily large magnitude contributing to the loop. Their contribution can only be suppressed by suppressing high frequencies, which can be achieved by considering sufficiently large time separations or integrating over sufficiently long times.

One way of extracting IR properties even in the equal-time limit would be to consider correlations between large spatial regions with no overlap. More specifically, suppose that we smear the two-point function with two different smearing functions $F_1(x)$ and $F_2(x)$ with no overlapping spatial support. The corresponding smeared correlation function will be finite even when $\sigma_t \to 0$. The relation to actual measurements, however, deserves further discussion. Let us consider observables with a certain spatial extension and time duration $\sigma_t$, characterized by smearing functions $F_1(x)$ and $F_2(x)$. One can think of the ensemble with the outcomes of measuring these observables if they were repeated many times. For very small $\sigma_t$ the statistical dispersion for the measurements of the observables associated with $F_1(x)$ or $F_2(x)$ would become very large even if their spatial support is large. On the other hand, the correlator of the two would be finite even in the limit $\sigma_t \to 0$ if the spatial supports of $F_1(x)$ and $F_2(x)$ do not overlap: each one exhibits large fluctuations, but they are mutually uncorrelated. Note, however, that the product of the two measurements will in general be very large for each realization and only the average over many realizations would agree with the quantum correlation function and have a moderate value.

Let us illustrate all this with a concrete example: the current measurement of CMB anisotropies. If one were to perform the measurements for sufficiently short times, loop corrections from QED would eventually give rise to large fluctuations, which would be due to UV effects of the current vacuum polarization, having nothing to do with IR effects of cosmological origin. Such large statistical fluctuations would go away, after averaging over the outcomes of a large number of repeated measurements, if one correlated the measurements corresponding to non-overlapping solid angles with non-vanishing angular size. Nevertheless, such spurious statistical fluctuations can only be avoided for single realizations if the measurement lasts for a minimum period of time.


\bibliography{literature}

\end{document}